\documentclass[Conference,a4paper]{IEEEtran}
\IEEEoverridecommandlockouts
\usepackage{color}
\usepackage{graphicx}
\usepackage{epstopdf}
\usepackage{amsmath}
\usepackage{amssymb}
\usepackage{algorithm}
\usepackage{algorithmic}
\usepackage{amsmath}
\usepackage{multirow}
\usepackage{booktabs}
\usepackage{array}
\usepackage{amsthm}
\usepackage{stfloats}
\usepackage{caption}
\usepackage{subfigure}
\usepackage{bm}
\usepackage{booktabs}
\usepackage{setspace}
\usepackage{diagbox}
\usepackage{enumerate}
\usepackage{ulem}
\usepackage{url}
\usepackage{circledsteps}
\usepackage{graphicx}
\usepackage{pifont}
\usepackage{makecell}
\newcommand{\cmark}{\ensuremath{\checkmark}}
\newcommand{\xmark}{\ensuremath{\times}}

\usepackage{hyperref}
\allowdisplaybreaks[4]

{\bgroup
 \addtolength\abovedisplayshortskip{#1}
 \addtolength\abovedisplayskip{#1}
 \addtolength\belowdisplayshortskip{#1}
 \addtolength\belowdisplayskip{#1}
 }
{\egroup\ignorespacesafterend}

\newtheorem{lemma}{Lemma}

\newcommand{\be}{\begin{equation}}
\newcommand{\ee}{\end{equation}}
\newcommand{\bea}{\begin{eqnarray}}
\newcommand{\eea}{\end{eqnarray}}
\newcommand{\ba}{\begin{array}}
\newcommand{\ea}{\end{array}}

\title{
Reconfigurable Holographic Surface for Simultaneous Wireless Information and Power Transfer
}
\author{\IEEEauthorblockN{Yuan Guo, Wen Chen, Ziwei Liu, Chaoying Huang, Zhendong Li, and Ying Wang
}
\thanks{ Y. Guo, 
W. Chen, 
Z. Liu,
and 
C. Huang are with the Department of Electronic Engineering, Shanghai Jiao Tong University, Shanghai, China, email: yuanguo26@sjtu.edu.cn, wenchen@sjtu.edu.cn, ziweiliu@sjtu.edu.cn, chaoyinghuang@sjtu.edu.cn.}
\thanks{Z. Li is with the School of Information and Communication Engineering, Xi'an Jiaotong University, Xi'an, China, email: lizhendong@xjtu.edu.cn.}
\thanks{ Y. Wang is with the State Key Laboratory of Networking and Switching Technology,  
Beijing University of Posts and Telecommunications, Beijing, China, email: wangying@bupt.edu.cn.}
}

\begin{document}
\maketitle
\pagestyle{empty}
\thispagestyle{empty}

\begin{abstract}
In this paper, we propose to use a novel reconfigurable holographic surface (RHS) 
to improve the performance of simultaneous wireless information and power transfer (SWIPT) 
by exploiting the additional spatial degrees of freedom (DoFs) enabled by the holographic interference principle. 
Specifically, 
we study an RHS-empowered SWIPT system in which the digital beamformer at the base station (BS) 
and the holographic beamformer at the RHS 
are jointly optimized to maximize the weighted sum-rate of information-decoding (ID) users 
while guaranteeing a minimum harvested energy requirement for each energy-harvesting (EH) user. 
Due to the non-convexity of the optimization problem, 
the weighted sum-rate maximization problem is challenging to solve. 
We first adopt the weighted minimum mean squared error (WMMSE) method
to transform the objective into a more tractable form. 
We then develop an iterative optimization framework, 
where the BS digital beamforming and the RHS holographic beamforming subproblems are solved 
via the majorization-minimization (MM) method. 
Since solving each subproblem can incur prohibitive complexity 
as the variable dimension increases, 
we further propose low-complexity solutions for the two subproblems 
based on the alternating direction method of multipliers (ADMM) methodology.
Numerical results validate the convergence behavior of the proposed algorithms 
and demonstrate that the RHS-aided BS achieves significant performance gains 
over a conventional fully-digital BS benchmark. 
Moreover, 
the proposed low-complexity algorithms substantially reduce computational complexity 
while maintaining nearly the same performance.

\end{abstract}


\maketitle
\section{Introduction}


With the number of Internet of Things (IoT) devices increasing explosively, 
resource allocation of both spectrum and energy is inevitably constrained, 
which becomes a bottleneck for the development of next-generation wireless networks \cite{ref_IoT_1}.
In this context,
radio frequency (RF) transmission enabled simultaneous wireless information and power transfer (SWIPT) \cite{ref_SWIPT_1}$-$\cite{ref_SWIPT_2}
has emerged as a promising spectrum- and energy-efficient solution 
for addressing the issue of limited resources.
Specifically, 
a base station (BS) transmits RF signals to two groups of users to simultaneously convey information and deliver wireless power over the same time-frequency resources.
Under the separated SWIPT architecture, 
the receivers are divided into two groups according to their functionalities, 
namely information-decoding (ID) users and energy-harvesting (EH) users. 
The ID users recover the transmitted information from the received RF signals, 
whereas the EH users exploit them for energy harvesting.



Despite its potential, 
separated SWIPT is often limited by unfavorable propagation conditions (i.e., severe path loss),
particularly when EH users are far from the BS.
As a result,
this leads to high BS transmit power for meeting the EH requirements.
Recently,
reconfigurable intelligent surfaces (RISs) \cite{ref_RIS_1}$-$\cite{ref_RIS_2} have been introduced as a cost-effective technique to reconfigure the wireless propagation environment.
An RIS is a planar array composed of a large number of nearly passive elements.
These elements are capable of adjusting the phase of the signals, 
allowing for the creation of a favorable signal propagation condition. 
By manipulating the phase of the incident signals, 
RIS can enhance the received signal quality, 
improving both the coverage and the reliability of wireless communications \cite{ref_RIS_1}$-$\cite{ref_RIS_3}.
Motivated by these merits, 
an increasing number of studies have investigated the deployment of RIS in the SWIPT system 
to enhance system performance, i.e., \cite{ref_RIS_SWIPT_1}$-$\cite{ref_RIS_SWIPT_7}.
Despite its benefits,
RIS is typically designed with half-wavelength inter-element spacing, 
thereby limiting the number of elements within a fixed aperture.
This aperture-limited element count constrains the achievable spatial degrees of freedom (DoFs).

To overcome this limitation, 
reconfigurable holographic surface (RHS) has emerged as an attractive alternative \cite{ref_RHS_1}$-$\cite{ref_RHS_2}, 
enabling sub-wavelength element spacing and thus accommodating a much denser array of units.
Specifically,
an RHS can be viewed as a leaky-wave metasurface antenna that mainly consists of feeds, 
a guided-wave structure (e.g., a parallel-plate waveguide), 
and sub-wavelength metamaterial unit cells.
The feeds, integrated in the bottom layer, generate a guided reference wave. 
This reference wave propagates within the parallel-plate waveguide along the aperture and is leaked into free space via the spatially modulated unit cells, forming the desired radiation pattern.
As a result, 
an RHS can more easily realize a large-aperture antenna array with fewer active RF components, 
which reduces cost and power consumption.

Given the above advantages of the RHS architecture,
numerous works have studied RHS-enabled wireless networks from different aspects, 
aiming to enhance overall system performance, e.g., \cite{ref_RHS_application_1}$-$\cite{ref_RHS_application_10}.
For example,
the authors of \cite{ref_RHS_application_1} proposed a novel holographic-pattern division multiple access (HDMA) 
and showed that the sum-rate achieved by zero-forcing (ZF) precoding approaches the asymptotic capacity of the HDMA system.
In \cite{ref_RHS_application_2}, 
the authors studied an RHS-enabled ultra-dense low-Earth-orbit (LEO) satellite communication network
and proposed a holographic beamforming design to tackle the sum-rate maximization problem.
The paper \cite{ref_RHS_application_3} investigated a holographic multi-input multi-output surface (HMIMOS)-empowered 
non-orthogonal multiple access (NOMA) wireless communication scheme 
under both single-cluster and multi-cluster NOMA transmission scenarios.
The work \cite{ref_RHS_application_4} focused on an HMIMO-aided multi-cell system 
and designed a low-complexity solution to maximize the weighted sum-rate 
under practical per-RF chain power constraints.
In \cite{ref_RHS_application_5}, 
the authors aimed to improve the wireless energy transfer (WET) performance of the WET system
by leveraging an RHS-enabled transmitter 
while considering a practical nonlinear power-amplifier (PA) model.
Since acquiring real-time channel state information (CSI) for beamforming in large-scale RHS is prohibitively complex, 
the authors of \cite{ref_RHS_application_6} proposed a one-shot multi-user beam training method for both near- and far-field scenarios.
The work \cite{ref_RHS_application_7} integrated the RHS into the integrated sensing and communication (ISAC) system
to enhance both communication and sensing performance with lower power consumption.
Furthermore, \cite{ref_RHS_application_8} proposed an RHS beamforming framework 
that maximizes a novel mutual information (MI)-based sensing metric in an RHS-enabled ISAC system, 
and derived an asymptotic lower bound on the MI.
In \cite{ref_RHS_application_9},
RHS and RIS are jointly deployed in a wideband ISAC system to mitigate severe propagation losses and attenuation,
thereby improving the reliability of communication links and the sensing performance.
The authors of \cite{ref_RHS_application_10} investigated an RHS-unmanned aerial vehicle (UAV)-based ISAC system 
assisted by an RIS,
leveraging the UAV's mobility and the RIS's reconfigurability for flexible beam control,
while using the RHS to further enhance the overall sensing and communication performance.

\subsection{Motivations and Contributions}

\begin{table*}[!t]
\centering
\caption{{Comparison of Existing Works}}
\label{tab_existing_works_comparison}
\begin{tabular}{|c|c|c|c|c|c|}
\hline
\text{References} 
& \text{RHS Architecture}
& \text{SWIPT} 
& \text{ID/EH Users Coexistence} 
& \text{Holographic Beamforming}  \\
\hline

\cite{ref_RIS_SWIPT_1}$-$\cite{ref_RIS_SWIPT_4}
& \xmark 
& \cmark 
& \cmark
& \xmark \\
\hline

\cite{ref_RIS_SWIPT_5}$-$\cite{ref_RIS_SWIPT_6}
& \xmark
& \cmark
& \cmark
& \xmark \\
\hline

\cite{ref_RIS_SWIPT_7}
& \xmark 
& \cmark 
& \xmark
& \xmark \\
\hline

\cite{ref_RHS_application_5}
& \cmark
& \xmark
& \xmark
& \cmark \\
\hline

\text{This paper}
& \cmark
& \cmark
& \cmark
& \cmark \\
\hline

\end{tabular}
\end{table*}

{
Although RHSs have been widely investigated for performance enhancement in various wireless scenarios,
e.g., \cite{ref_RHS_application_1}$-$\cite{ref_RHS_application_10},
their application to SWIPT systems has received comparatively limited attention.
More importantly, 
as summarized in Table~\ref{tab_existing_works_comparison},
the joint digital and RHS holographic beamforming design for RHS-assisted SWIPT systems with coexisting ID and EH users has not been systematically studied.
Motivated by this observation, 
we investigate an RHS-assisted SWIPT system, 
where the RHS serves as a holographic transmit aperture to enable highly directional wavefront shaping for both information transmission and energy focusing.
Specifically,
we propose a hybrid beamforming framework that jointly optimizes the BS digital beamformer and the RHS holographic beamformer 
to maximize the sum-rate of ID users 
while satisfying the individual harvested-energy requirements of EH users under the real-domain amplitude constraints of the RHS.
The main contributions of this paper are summarized as follows:
}

\begin{itemize}
\item This paper investigates an RHS-assisted SWIPT system with multiple ID users and multiple EH users, aiming to simultaneously enhance communication and energy transfer. 
    We formulate an optimization problem to maximize the sum-rate of the ID users, subject to a minimum harvested-energy threshold for each EH user and the amplitude constraints of the RHS.
    The resulting problem is tackled via a joint design of the BS digital beamformer and the RHS holographic beamformer.
    Notably, the per-element real-valued amplitude constraints introduce a large number of constraints, which significantly complicates the beamforming design.

\item To solve the non-convex optimization problem with coupled variables (i.e., the digital and holographic beamformers),
    we develop an optimization algorithm that updates the two beamformers iteratively. 
    In particular, the sum-rate objective is first recast into an equivalent form via the weighted minimum mean squared error (WMMSE) method \cite{ref_WMMSE}. 
    Furthermore, by leveraging the majorization-minimization (MM) framework \cite{ref_MM},
    the resulting subproblems can be reformulated as second-order cone programs (SOCPs).

\item Moreover, since a large number of RF chains and/or RHS elements can significantly increase the computational complexity, we further propose low-complexity solutions for the digital and holographic beamformers, respectively, based on the alternating direction method of multipliers (ADMM) framework \cite{ref_ADMM}.

\item Last but not least, extensive numerical results demonstrate the advantages of the RHS architecture and the proposed algorithms.
    In particular, the results show that (i) RHS-assisted transmission significantly improves system performance compared with a conventional phased-array architecture, 
    and (ii) the proposed low-complexity solutions substantially reduce the computational complexity compared with SOCP-based benchmarks.

\end{itemize}

{
A list of the main notations is provided in Table \ref{tab:notation}.}

\begin{table*}[!t]
\caption{{Main Notations}}
\label{tab:notation}
\centering
\begin{tabular}{|c|c|c|c|}
\hline
\textbf{Notation} 
& \textbf{Description} 
& \textbf{Notation} 
& \textbf{Description} \\
\hline

$K$ & Number of ID users 
& $G$ & Number of EH users \\
\hline

$L$ & Number of RF chains at the BS 
& $N$ & Number of RHS elements \\
\hline

$\mathcal{K}$ & Set of ID users, i.e., $\mathcal{K}\triangleq \{1,\ldots,K\}$ 
& $\mathcal{G}$ & Set of EH users, i.e., $\mathcal{G}\triangleq \{1,\ldots,G\}$ \\
\hline

$\mathcal{L}$ & Set of RF chains, i.e., $\mathcal{L}\triangleq \{1,\ldots,L\}$ 
& $\mathcal{N}$ & Set of RHS elements, i.e., $\mathcal{N}\triangleq \{1,\ldots,N\}$ \\
\hline

$\mathbf{w}_{I,k}$ & Digital beamforming vector for  ID user $k$ 
& $\mathbf{w}_{E,g}$ & Energy beamforming vector for  EH user $g$ \\
\hline

$\boldsymbol{\Psi}$ & RHS holographic beamforming matrix 
& $\psi_n$ & Amplitude coefficient of RHS unit $n$ \\
\hline

$\mathbf{h}_{I,k}$ & Channel vector from the RHS to ID user $k$ 
& $\mathbf{h}_{E,g}$ & Channel vector from the RHS to EH user $g$ \\
\hline

$P_t$ & Total transmit power budget 
& $Q_t$ & Minimum harvested energy requirement \\
\hline

\end{tabular}
\end{table*}

\section{System Model and Problem Formulation}

\subsection{RHS Holographic Interference Principle}
\begin{figure}[t]
	\centering
	\includegraphics[width=.40\textwidth]{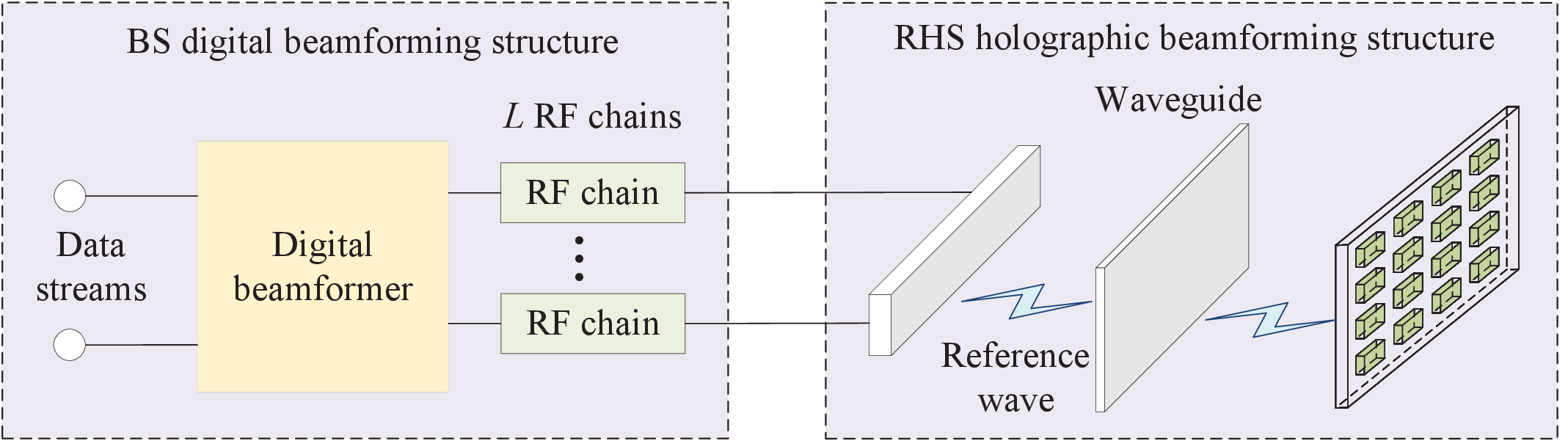}
	\caption{The block diagram of the digital and holographic beamforming.}
	\label{fig.2}
\end{figure}

{
In this subsection, 
we briefly introduce the holographic interference principle of the RHS \cite{ref_RHS_1}$-$\cite{ref_RHS_2}, 
which provides the physical basis for the real-valued amplitude-control model adopted in this paper.
As shown in Fig.~\ref{fig.2}, 
an RHS is a leaky-wave metasurface antenna mainly consisting of $L$ feeds, 
a guided-wave structure, 
and $N$ sub-wavelength metamaterial radiating elements.
The feeds are connected to the RF chains and generate guided reference waves carrying the transmitted signals. 
These reference waves propagate along the waveguide and sequentially excite the radiating elements. 
Moreover,
since each RF chain is connected to one feed of the RHS, 
the number of RF chains is the same as the number of feeds, i.e., $L$, 
and there is no channel or attenuation between the BS and the RHS \cite{ref_RHS_1}.
During this propagation process, 
part of the guided-wave energy is leaked into free space through the radiating elements, 
forming the radiated wavefront.
The sets of  feeds (RF chains), and the RHS's radiation units are defined as
$\mathcal{L} \triangleq \{1,\cdots,L\} $
and $\mathcal{N} \triangleq \{1,\cdots,N\} $,
respectively.

The key idea of holographic beamforming is to construct a radiation-amplitude pattern on the RHS according to the interference between the guided reference wave and the desired object wave.
For the $l$-th feed and the $n$-th radiating element, 
the reference wave can be expressed as
$\Psi_{ ref}(\mathbf{r}_{n}^{l})
=
\exp(-j\mathbf{k}_{s}\cdot\mathbf{r}_{n}^{l})$,
where $\mathbf{k}_{s}$ denotes the propagation vector of the guided 
reference wave, and $\mathbf{r}_{n}^{l}$ is the distance vector from 
the $l$-th feed to the $n$-th radiating element. To generate a desired 
radiation beam towards a given direction, the corresponding object 
wave is given by
$\Psi_{ obj}(\mathbf{r}_{n})
=
\exp(-j\mathbf{k}_{f}\cdot\mathbf{r}_{n})$,
where $\mathbf{k}_{f}$ is the propagation vector of the desired 
free-space wave and $\mathbf{r}_{n}$ denotes the position vector of 
the $n$-th radiating element.

According to the holographic principle, the interference pattern 
between the reference wave and the object wave is written as
$\Psi_{ intf}(\mathbf{r}_{n}^{l})
=
\Psi_{ obj}(\mathbf{r}_{n})
\Psi_{ ref}^{*}(\mathbf{r}_{n}^{l})$.
When this interference pattern is excited by the reference wave, the 
desired object wave can be reconstructed since
$\Psi_{ intf}(\mathbf{r}_{n}^{l})\Psi_{\rm ref}(\mathbf{r}_{n}^{l})
\propto
\Psi_{ obj}(\mathbf{r}_{n})
|\Psi_{ ref}(\mathbf{r}_{n}^{l})|^2$.

In practical RHS implementations, the radiating elements mainly 
control the leakage amplitude of the reference wave rather than 
providing independent phase shifts. Therefore, the normalized 
radiation amplitude of the $n$-th element can be modeled as a 
real-valued coefficient
$\psi_n =
\frac{\text{Re}\{\Psi_{ intf}(\mathbf{r}_{n}^{l})\}+1}{2}$,
which satisfies
$0 \leq \psi_n \leq 1$, 
$\forall n \in \mathcal{N}$.

The above holographic principle can be extended to the multi-beam case through holographic-pattern superposition,
which is also known as HDMA \cite{ref_RHS_application_1}. 
Suppose that the RHS is expected to generate multiple desired beams towards different
directions indexed by $u\in\mathcal{U}$. 
The corresponding normalized holographic pattern can be written as a weighted superposition of
the single-beam holographic patterns, i.e.,
$m_n
=
\sum_{u\in\mathcal{U}}
\sum_{l\in\mathcal{L}}
a_{u,l}
m(\mathbf{r}_{n}^{l},\theta_u,\varphi_u)$,
where $m(\mathbf{r}_{n}^{l},\theta_u,\varphi_u)$ denotes the
holographic pattern associated with the $u$-th desired direction and
the $l$-th feed, and $a_{u,l}$ is the corresponding amplitude ratio.
With proper normalization, e.g.,
$\sum_{u\in\mathcal{U}}\sum_{l\in\mathcal{L}}a_{u,l}=1$,
the resulting holographic pattern still satisfies
$0\leq m_n\leq 1$,
$\forall n\in\mathcal{N}$.
}

\subsection{System Model}

\begin{figure}[t]
	\centering
	\includegraphics[width=.30\textwidth]{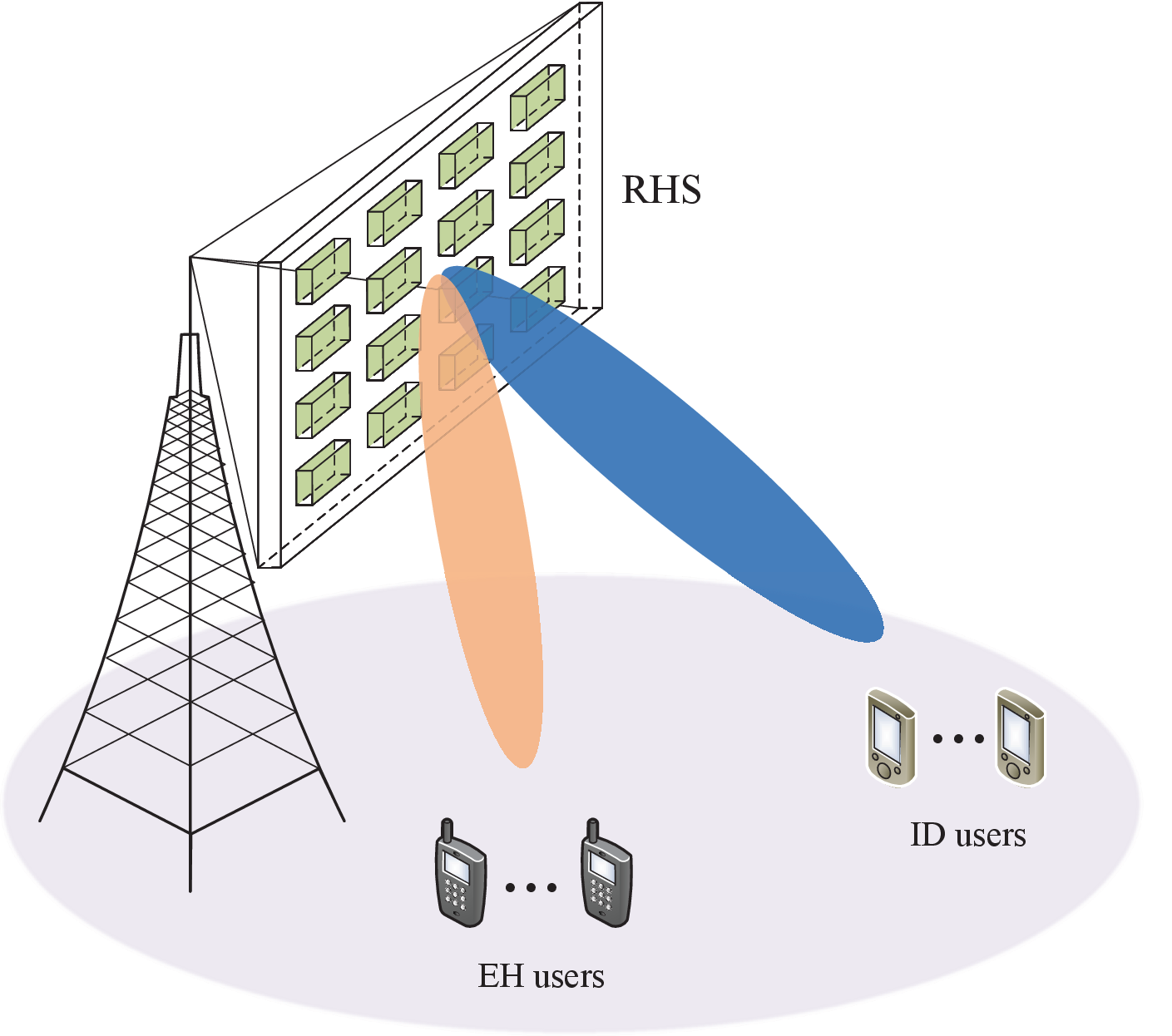}
	\caption{{ An illustration of the RHS-enabled SWIPT system.}}
	\label{fig.1}
\end{figure}

Fig. \ref{fig.1}  presents the considered RHS-aided SWIPT system,
where a BS equipped with one RHS  simultaneously serves $K$ single-antenna ID users and $G$ single-antenna EH users.
Besides,
the sets of ID users and EH users are defined as
$\mathcal{K} \triangleq \{1,\cdots,K\} $
and
$\mathcal{G} \triangleq \{1,\cdots,G\} $,
respectively.

First,
the signal transmitted from the BS via $L$ RF chains is expressed as
\begin{align}
\mathbf{x} = {\sum}_{k=1}^{K}\mathbf{w}_{I,k}s_{I,k}
+{\sum}_{g=1}^{G}\mathbf{w}_{E,g}s_{E,g}, \label{signal_model_1}
\end{align}
where
$s_{I,k}$
and 
$s_{E,g}$
are the communication and energy signals, respectively.
Without loss of generality,
we assume that 
$s_{I,k}$
and 
$s_{E,g}$
follow an independent and identically distributed (i.i.d.) circularly symmetric complex Gaussian distribution with zero mean and unit variance, i.e., $\mathcal{CN}(0,1)$.
$\mathbf{w}_{I,k}\in \mathbb{C}^{L\times 1}$
and 
$\mathbf{w}_{E,g}\in \mathbb{C}^{L\times 1}$
denote the digital beamforming vectors for the $k$-th ID user and the $g$-th EH user,
respectively.

Furthermore,
following the operating principle of the RHS \cite{ref_RHS_1}$-$\cite{ref_RHS_2},
the radiated signal can be expressed as
\begin{align}
\mathbf{x}_{RHS} = \boldsymbol{\Psi}\mathbf{Q}\mathbf{x},\label{signal_model_2}
\end{align}
where
$\boldsymbol{\Psi} = \text{diag}(\boldsymbol{\psi}) \in \mathbb{R}^{N\times N} $ 
is the holographic beamforming matrix of the RHS,
and
the vector $\boldsymbol{\psi}=[\psi_1,\cdots,\psi_N]^T\in\mathbb{R}^{N\times 1}$ collects the amplitude coefficients of the $N$ radiating elements, with $0\le \psi_n\le 1$.
Moreover, 
$\mathbf{Q}\in\mathbb{C}^{N\times L}$ denotes the phase shift matrix from the $L$ feeds to the $N$ radiating units,
in which the $(n,l)$-th element of $\mathbf{Q}$ is $[\mathbf{Q}]_{n,l}=e^{-j2\pi\kappa d_{n,l} /{\lambda}}$,
where
${\lambda}$ denotes the wavelength of the reference wave, 
$d_{n,l}$ is the distance from the $l$-th feed to the $n$-th unit,
and $\kappa$ represents the refractive index.

{
In this work, we assume that the CSI can be perfectly estimated, 
which provides a performance benchmark for the proposed RHS-enabled SWIPT beamforming design.
\footnote{{
The robust beamforming design under imperfect CSI is an important and practically relevant research direction for RHS-enabled SWIPT systems, which will be investigated in our future work.}
}}
The signal received by the $k$-th ID user can be given by
\begin{align}
{y}_{I,k} &= 
{\sum}_{i=1}^{K} \mathbf{h}_{I,k}^H\mathbf{C}_{M}\boldsymbol{\Psi}\mathbf{Q}\mathbf{w}_{I,i}s_{I,i}\label{Communication_signal}\\
&+{\sum}_{g=1}^{G} \mathbf{h}_{I,k}^H\mathbf{C}_{M}\boldsymbol{\Psi}\mathbf{Q}\mathbf{w}_{E,g}s_{E,g}
+ n_{I,k},\nonumber
\end{align}
where
the term 
$n_{I,k}\sim \mathcal{CN}(0, \sigma_{I,k}^2)$ 
denotes  additive white Gaussian noise (AWGN).
$\mathbf{h}_{I,k} \in \mathbb{C}^{N\times 1}$ is the wireless channel from the RHS to the $k$-th ID user
in the absence of mutual coupling.
$\mathbf{C}_{M} \in \mathbb{C}^{N\times N}$ represents the mutual coupling matrix caused by the dense arrangement of RHS units \cite{ref_mutula_coupling}$-$\cite{ref_mutula_coupling_1}.
Specifically,
the matrix $\mathbf{C}_{M}$ can be expressed as
\begin{align}
\mathbf{C}_{M}
=(\mathbf{I}_N+ \mathbf{Z}_{O}\mathbf{Z}_{T}^{-1} )^{-1}
(\mathbf{I}_N+ \mathbf{Z}_{O}\mathbf{Z}_{A,T}^{-1}),
\end{align}
where
$\mathbf{Z}_{O} = {Z}_{O}\mathbf{I}_{N}$ 
is the output impedance matrix of the voltage sources,
and 
$\mathbf{Z}_{T}$ denotes the mutual impedance matrix.
When mutual coupling is absent,
the matrix $\mathbf{Z}_{T}$ reduces to the antenna impedance matrix
$\mathbf{Z}_{A,T} = {Z}_{A,T}\mathbf{I}_{N}$ with the antenna impedance parameter ${Z}_{A,T}$.
Moreover,
the diagonal elements of the matrix $\mathbf{Z}_{T}$ are equal to ${Z}_{A,T}$,
and the $(p,q)$-th $(p\neq q)$ unit of the matrix $\mathbf{Z}_{T}$
can be obtained by (\ref{mutual_Z}),
\begin{figure*}
\begin{small}
\begin{align}
[\mathbf{Z}_{T}]_{p,q}
\!=\! -30 \left\{
\int_{0}^{l_d} \sin\!\left(\frac{2\pi}{\lambda} z\right)
\!+\! \int_{l_d}^{2l_d} \sin\!\left[\frac{2\pi}{\lambda}\left(2l_d\!-\!z\right)\right]
\right\}
\left(
\frac{-j e^{-j\frac{2\pi}{\lambda} r_1}}{r_1}
\!+\! \frac{-j e^{-j\frac{2\pi}{\lambda} r_2}}{r_2}
\!+\! \frac{2j \cos\!\left(\frac{2\pi}{\lambda} l_d\right)e^{-j\frac{2\pi}{\lambda} r_0}}{r_0}
\right)\, dz,\label{mutual_Z}
\end{align}
\end{small}
\boldsymbol{\hrule}
\end{figure*}
where
the variables
$r_0$, $r_1$, and $r_2$ are respectively defined as
\begin{align}
&r_0 = \sqrt{d_{p,q}^2 + z^2}, 
r_1 = \sqrt{d_{p,q}^2 + (l_d - z)^2}, \nonumber\\
&r_2 = \sqrt{d_{p,q}^2 + (l_d + z)^2},\label{mutual_r}
\end{align}
in which $d_{p,q}$ is the spacing between the $p$-th and $q$-th elements,
and $l_d$ denotes the antenna length.

Then,
the received signal at EH user $g$ can be written as
\begin{align}
{y}_{E,g} &= 
{\sum}_{i=1}^{G} \mathbf{h}_{E,g}^H\mathbf{C}_{M}\boldsymbol{\Psi}\mathbf{Q}\mathbf{w}_{E,i}s_{E,i}\\
&+{\sum}_{k=1}^{K} \mathbf{h}_{E,g}^H\mathbf{C}_{M}\boldsymbol{\Psi}\mathbf{Q}\mathbf{w}_{I,k}s_{I,k}
+ n_{E,g},\nonumber
\end{align}
where 
$n_{E,g}\sim \mathcal{CN}(0, \sigma_{E,g}^2)$ is the AWGN
and
$\mathbf{h}_{E,g}$ denotes the channel from the RHS to the $g$-th EH user in the absence of mutual coupling.

{  
Note that the mutual coupling matrix $\mathbf C_M$ is determined by the RHS geometry and the impedance parameters, 
and is treated as a known matrix in the beamforming design.  
It is absorbed into the effective channel 
$\mathbf h^H\mathbf C_M\boldsymbol{\Psi}\mathbf Q$, 
thereby changing the effective channel gain and the spatial correlation among RHS radiation units, 
but not the basic input-output structure of the considered RHS-enabled SWIPT system.  
Moreover, 
since $\mathbf C_M$ is not an optimization variable, 
it does not alter the applicability of the algorithms developed in Section III.
}

\subsection{Performance Metrics}

\textit{1) Communication metric:}
Based on the received signal (\ref{Communication_signal}),
the SINR of the $k$-th ID user is given in (\ref{SINR}).
\begin{figure*}
\begin{align}
\text{SINR}_{k}( \{\mathbf{w}_{I,k}\}, \{\mathbf{w}_{E,g}\}, \boldsymbol{\Psi} )
=
\frac{\vert\mathbf{h}_{I,k}^H\mathbf{C}_{M}\boldsymbol{\Psi}\mathbf{Q}\mathbf{w}_{I,k}  \vert^2}
{
\sum_{i\neq k}^{K}\vert \mathbf{h}_{I,k}^H\mathbf{C}_{M}\boldsymbol{\Psi}\mathbf{Q}\mathbf{w}_{I,i}\vert^2
+
\sum_{g=1}^{G} \vert \mathbf{h}_{I,k}^H\mathbf{C}_{M}\boldsymbol{\Psi}\mathbf{Q}\mathbf{w}_{E,g}\vert^2
+
\sigma_{I,k}^2
}.
\label{SINR}
\end{align}
\boldsymbol{\hrule}
\end{figure*}
Thus,
the achievable rate of ID user $k$ can be formulated as
\begin{align}
\mathrm{R}_{k}( \{\mathbf{w}_{I,k}\}, \{\mathbf{w}_{E,g}\}, \boldsymbol{\Psi} ) =
\log( 1 +\text{SINR}_{k}).
\end{align}

\textit{2) Energy harvesting metric:}
First,
to evaluate the upper bound of the energy harvesting performance,
we assume that the EH users adopt a linear EH model \cite{ref_EH_model}.
\footnote{{
Practical nonlinear EH modeling for RHS-enabled SWIPT systems is an important research direction and will be investigated in our future work.}
}
Therefore,
the harvested energy at EH user $g$ is given by
\begin{align}
&\mathrm{Q}_{g}( \{\mathbf{w}_{I,k}\}, \{\mathbf{w}_{E,g}\}, \boldsymbol{\Psi}  )\\
&=\zeta
\bigg(
\sum_{i=1}^{G} \vert\mathbf{h}_{E,g}^H\mathbf{C}_{M}\boldsymbol{\Psi}\mathbf{Q}\mathbf{w}_{E,i}\vert^2
\!+\!\!
\sum_{k=1}^{K} \vert\mathbf{h}_{E,g}^H\mathbf{C}_{M}\boldsymbol{\Psi}\mathbf{Q}\mathbf{w}_{I,k}\vert^2
\bigg)
,\nonumber
\end{align}
where
$0< \zeta \leq 1$ denotes the energy harvesting efficiency.

\subsection{Problem Formulation}

In this work,
we aim to maximize the weighted sum-rate of all ID users
while guaranteeing the energy harvesting demands of all EH users
by jointly optimizing the BS digital beamforming $(\{\mathbf{w}_{I,k}\}, \{\mathbf{w}_{E,g}\})$ 
and the RHS holographic beamforming $\boldsymbol{\Psi}$.
Based on the above, 
the corresponding optimization problem can be formulated as
\begin{subequations}
\begin{align}
\textrm{(P0)}:&\mathop{\textrm{max}}
\limits_{\{\mathbf{w}_{I,k}\}, \{\mathbf{w}_{E,g}\}, \boldsymbol{\Psi}
}\
{\sum}_{k=1}^{K}
\upsilon_{I,k}
\mathrm{R}_{k}(\{\mathbf{w}_{I,k}\}, \{\mathbf{w}_{E,g}\}, \boldsymbol{\Psi}) 
\label{P0_obj}\\
\textrm{s.t.}\ 
&\mathrm{Q}_g(\{\mathbf{w}_{I,k}\}, \{\mathbf{w}_{E,g}\}, \boldsymbol{\Psi})  \geq Q_t, \forall g \in \mathcal{G}, \label{P0_c_1}\\
&{\sum}_{k=1}^{K}\mathbf{w}_{I,k}^H\mathbf{w}_{I,k}
+{\sum}_{g=1}^{G}\mathbf{w}_{E,g}^H\mathbf{w}_{E,g}\leq P_t,\label{P0_c_2}\\
& 0 \leq [\boldsymbol{\psi}]_n \leq 1, \forall n \in \mathcal{N}, \label{P0_c_3}
\end{align}
\end{subequations}
where
$\upsilon_{I,k}$ is the weighting coefficient of ID user $k$,\footnote{{
Since equal-priority ID users are considered in this paper, 
we set $\upsilon_{I,k}=1$, $\forall k$, 
which does not affect the subsequent algorithm design.
}
}
$Q_t$ denotes the minimum EH threshold,
and
$P_t$ represents the maximum transmit power of the BS.
It is observed that problem (P0) is highly challenging due to  the non-convexity of both the objective and the constraints.

\section{Algorithm Design}

\subsection{Problem Reformulation}

Note that the sum-rate objective involves fractional terms.
The resulting problem is highly non-convex and cannot be handled by standard convex optimization tools.
Therefore,
we adopt the WMMSE method \cite{ref_WMMSE}  to transform the sum-rate 
into an equivalent form by introducing auxiliary variables $\{\beta_k \}$ and $\{\omega_k \}$,
which can be given in (\ref{WMMSE_transformation}). \footnote{{
For fixed holographic and digital beamformers, 
the received signal at each ID user has the standard form required by the WMMSE method \cite{ref_WMMSE}. 
The RHS-related matrices $\mathbf Q$, $\mathbf C_M$, and $\boldsymbol{\Psi}$ are absorbed into the effective channel 
$\mathbf h_{I,k}^{H}\mathbf C_M\boldsymbol{\Psi}\mathbf Q$,
and thus do not change the input-output structure.
Therefore, 
the WMMSE transformation remains valid for the considered RHS-enabled SWIPT model.}}
\begin{figure*}
\begin{small}
\begin{align}
&\mathrm{R}_{k}( \{\mathbf{w}_{I,k}\}, \{\mathbf{w}_{E,g}\}, \boldsymbol{\Psi} ) 
=
\log
\bigg(
1
+
\frac{\vert\mathbf{h}_{I,k}^H\mathbf{C}_{M}\boldsymbol{\Psi}\mathbf{Q}\mathbf{w}_{I,k}  \vert^2}
{
\sum_{i\neq k}^{K}\vert \mathbf{h}_{I,k}^H\mathbf{C}_{M}\boldsymbol{\Psi}\mathbf{Q}\mathbf{w}_{I,i}\vert^2
+
\sum_{g=1}^{G} \vert \mathbf{h}_{I,k}^H\mathbf{C}_{M}\boldsymbol{\Psi}\mathbf{Q}\mathbf{w}_{E,g}\vert^2
+
\sigma_{I,k}^2
}
\bigg)\label{WMMSE_transformation}
\\
&= 
\mathop{\textrm{max}}
\limits_{
\omega_k\geq0,
\beta_k
}
\bigg\{
\underbrace{
\log(\omega_k)
\!-\!\omega_k\bigg( \!
1\!-\!2\text{Re}\{ \beta_k^{\ast} \mathbf{h}_{I,k}^H\mathbf{C}_{M}\boldsymbol{\Psi}\mathbf{Q}\mathbf{w}_{I,k}\}
\!\!+\! \vert\beta_k\vert^2(
{\sum}_{i=1}^{K}\vert \mathbf{h}_{I,k}^H\mathbf{C}_{M}\boldsymbol{\Psi}\mathbf{Q}\mathbf{w}_{I,i}\vert^2
\!\!+\!\!
{\sum}_{g=1}^{G} \vert \mathbf{h}_{I,k}^H\mathbf{C}_{M}\boldsymbol{\Psi}\mathbf{Q}\mathbf{w}_{E,g}\vert^2
\!+\!
\sigma_{I,k}^2)
\bigg) \!+\! 1}
\limits_{\mathrm{\tilde{R}}_k(\{\mathbf{w}_{I,k}\}, \{\mathbf{w}_{E,g}\}, \boldsymbol{\Psi},\omega_k,\beta_k)} \bigg\}.\nonumber
\end{align}
\end{small}
\boldsymbol{\hrule}
\end{figure*}
Therefore,
(P0) can be rewritten as
\begin{subequations}
\begin{align}
\textrm{(P1)}:&\mathop{\textrm{max}}
\limits_{\{\mathbf{w}_{I,k}\}, \{\mathbf{w}_{E,g}\}, \boldsymbol{\Psi},
\{\omega_k\},\{\beta_k\}
}\
{\sum}_{k=1}^{K}
\mathrm{\tilde{R}}_{k}
\label{P1_obj}\\
\textrm{s.t.}\ 
&\mathrm{Q}_g(\{\mathbf{w}_{I,k}\}, \{\mathbf{w}_{E,g}\}, \boldsymbol{\Psi})  \geq Q_t, \forall g \in \mathcal{G}, \label{P1_c_1}\\
&{\sum}_{k=1}^{K}\mathbf{w}_{I,k}^H\mathbf{w}_{I,k}
+{\sum}_{g=1}^{G}\mathbf{w}_{E,g}^H\mathbf{w}_{E,g}\leq P_t,\label{P1_c_2}\\
& 0 \leq [\boldsymbol{\psi}]_n \leq 1, \forall n \in \mathcal{N}. \label{P1_c_3}
\end{align}
\end{subequations}

Note that the optimization variables are highly coupled in the objective (\ref{P1_obj}) and constraints (\ref{P1_c_1}).
Next,
we adopt the block coordinate ascent (BCA) method \cite{ref_BCA} to efficiently solve problem (P1).

\subsection{Optimizing auxiliary variables}

According to the derivation of the WMMSE method \cite{ref_WMMSE},
when other variables are fixed,
the optimal solutions of the auxiliary variables can be obtained analytically as follows (\ref{beta_opt})$-$(\ref{omega_opt}).
\begin{figure*}
\begin{align}
&\beta_k^{\star}
=
\frac{\mathbf{h}_{I,k}^H\mathbf{C}_{M}\boldsymbol{\Psi}\mathbf{Q}\mathbf{w}_{I,k}}
{\sum_{i=1}^{K}\vert \mathbf{h}_{I,k}^H\mathbf{C}_{M}\boldsymbol{\Psi}\mathbf{Q}\mathbf{w}_{I,i}\vert^2
+
\sum_{g=1}^{G} \vert \mathbf{h}_{I,k}^H\mathbf{C}_{M}\boldsymbol{\Psi}\mathbf{Q}\mathbf{w}_{E,g}\vert^2
+
\sigma_{I,k}^2}, \label{beta_opt}\\
&\omega_k^{\star}
=
1+
\frac{\vert\mathbf{h}_{I,k}^H\mathbf{C}_{M}\boldsymbol{\Psi}\mathbf{Q}\mathbf{w}_{I,k}\vert^2}
{\sum_{i\neq k}^{K}\vert \mathbf{h}_{I,k}^H\mathbf{C}_{M}\boldsymbol{\Psi}\mathbf{Q}\mathbf{w}_{I,i}\vert^2
+
\sum_{g=1}^{G} \vert \mathbf{h}_{I,k}^H\mathbf{C}_{M}\boldsymbol{\Psi}\mathbf{Q}\mathbf{w}_{E,g}\vert^2
+
\sigma_{I,k}^2}.\label{omega_opt}
\end{align}
\boldsymbol{\hrule}
\end{figure*}

\subsection{Updating The Digital Beamformer}
In this subsection, 
we study the optimization of the digital beamforming $(\{\mathbf{w}_{I,k}\}, \{\mathbf{w}_{E,g}\})$.
First,
we rewrite the function $\mathrm{\tilde{R}}_{k}$ of (P1) as follows
\begin{align}
&\mathrm{\tilde{R}}_{k}
\Leftrightarrow
-{\sum}_{i=1}^{K}\mathbf{w}_{I,i}^H\mathbf{B}_{1,k}\mathbf{w}_{I,i}\\
&-{\sum}_{g=1}^{G}\mathbf{w}_{E,g}^H\mathbf{B}_{1,k}\mathbf{w}_{E,g}
+2\text{Re}\{ \mathbf{b}_{1,k}^H\mathbf{w}_{I,k} \} +c_{1,k},\nonumber
\end{align}
where
the new notations are given as
\begin{align}
&\mathbf{B}_{1,k} \triangleq
\omega_k\vert\beta_k\vert^2(\mathbf{Q}^H\boldsymbol{\Psi}^H\mathbf{C}_{M}^H\mathbf{h}_{I,k} \mathbf{h}_{I,k}^H\mathbf{C}_{M}\boldsymbol{\Psi}\mathbf{Q}),\\
&c_{1,k}
\triangleq\log(\omega_k) - \omega_k -\omega_k\vert\beta_k\vert^2\sigma_{I,k}^2+1,\nonumber\\
&\mathbf{b}_{1,k}\triangleq \omega_k\beta_k \mathbf{Q}^H\boldsymbol{\Psi}^H\mathbf{C}_{M}^H\mathbf{h}_{I,k}.\nonumber
\end{align}

And then,
the objective of (P1) can be reformulated as
\begin{align}
{\sum}_{k=1}^{K}\mathrm{\tilde{R}}_{k}
\Leftrightarrow
&- {\sum}_{k=1}^{K} \mathbf{w}_{I,k}^H\bigg({\sum}_{i=1}^{K}\mathbf{B}_{1,i}\bigg)\mathbf{w}_{I,k}\\
&- {\sum}_{g=1}^{G} \mathbf{w}_{E,g}^H\bigg({\sum}_{i=1}^{K}\mathbf{B}_{1,i}\bigg)\mathbf{w}_{E,g}
\nonumber  \\
& + {\sum}_{k=1}^{K} 2\text{Re}\{ \mathbf{b}_{1,k}^H\mathbf{w}_{I,k} \} + {\sum}_{k=1}^{K}c_{1,k}  \nonumber \\
&=-\mathbf{w}_{IE}^H\mathbf{B}_5\mathbf{w}_{IE}+2\text{Re}\{ \mathbf{b}_3^H\mathbf{w}_{IE} \} +c_2,\nonumber 
\end{align}
where
\begin{align}
&\mathbf{w}_I\triangleq
[\mathbf{w}_{I,1}^T,\cdots,\mathbf{w}_{I,K}^T]^T\in \mathbb{C}^{LK\times 1},\nonumber\\
&\mathbf{w}_E\triangleq
[\mathbf{w}_{E,1}^T,\cdots,\mathbf{w}_{E,G}^T]^T\in \mathbb{C}^{LG\times 1},\nonumber\\
&\mathbf{B}_{3} \triangleq \mathbf{I}_K \otimes\big({\sum}_{k=1}^{K}  \mathbf{B}_{1,k}\big),
\mathbf{B}_{4} \triangleq \mathbf{I}_G \otimes\big({\sum}_{k=1}^{K}  \mathbf{B}_{1,k}\big),\nonumber\\
&\mathbf{b}_{2}\triangleq [\mathbf{b}_{1,1}^T,\cdots,\mathbf{b}_{1,K}^T]^T,
c_2 = {\sum}_{k=1}^{K} c_{1,k},\nonumber\\
&\mathbf{w}_{IE}\triangleq [\mathbf{w}_I^T,\mathbf{w}_E^T]^T,
\mathbf{B}_5 \triangleq \text{blkdiag}(\mathbf{B}_{3},\mathbf{B}_{4}),
\mathbf{b}_{3}\triangleq [\mathbf{b}_{2}^T, \mathbf{0}_{G}^T]^T.\nonumber
\end{align}

Furthermore,
the constraints (\ref{P1_c_1}) and (\ref{P1_c_2})  of (P1) can be rewritten as
\begin{align}
&\mathrm{Q}_g(\{\mathbf{w}_{I,k}\}, \{\mathbf{w}_{E,g}\}, \boldsymbol{\Psi})  \geq Q_t
\Leftrightarrow 
\mathbf{w}_{IE}^H\mathbf{B}_{8,g}\mathbf{w}_{IE}\geq Q_t,\\
& {\sum}_{k=1}^{K}\mathbf{w}_{I,k}^H\mathbf{w}_{I,k}
+{\sum}_{g=1}^{G}\mathbf{w}_{E,g}^H\mathbf{w}_{E,g}\leq P_t\\
&\Leftrightarrow \mathbf{w}_{IE}^H\mathbf{w}_{IE}\leq P_t,\nonumber
\end{align}
respectively,
where the above newly introduced coefficients are formulated as
\begin{align}
&\mathbf{B}_{2,g} \triangleq
\mathbf{Q}^H\boldsymbol{\Psi}^H\mathbf{C}_{M}^H\mathbf{h}_{E,g} \mathbf{h}_{E,g}^H\mathbf{C}_{M}\boldsymbol{\Psi}\mathbf{Q},
\mathbf{B}_{6,g} \triangleq \mathbf{I}_G \otimes \mathbf{B}_{2,g},\nonumber\\
&\mathbf{B}_{7,g} \triangleq \mathbf{I}_K \otimes \mathbf{B}_{2,g},
\mathbf{B}_{8,g} \triangleq \text{blkdiag}(\mathbf{B}_{7,g},\mathbf{B}_{6,g}).
\end{align}

Then the problem (P1) can be equivalently formulated as
\begin{subequations}
\begin{align}
\textrm{(P2)}:&\mathop{\textrm{max}}
\limits_{\mathbf{w}_{IE}
}\
-\mathbf{w}_{IE}^H\mathbf{B}_5\mathbf{w}_{IE}+2\text{Re}\{ \mathbf{b}_3^H\mathbf{w}_{IE} \} + c_2
\label{P2_obj}\\
\textrm{s.t.}\ 
&\mathbf{w}_{IE}^H \mathbf{B}_{8,g}\mathbf{w}_{IE}  \geq Q_t, \forall g \in \mathcal{G}, \label{P2_c_1}\\
&\mathbf{w}_{IE}^H\mathbf{w}_{IE}\leq P_t.\label{P2_c_2}
\end{align}
\end{subequations}

Note that problem (P2) is intractable due to the non-convex constraints (\ref{P2_c_1}).
We apply the MM method to linearize constraints (\ref{P2_c_1}), which yields tight lower bounds.
Specifically,
following the MM framework \cite{ref_MM}, \cite{ref_MM_1},
we can linearize the left-hand side of (\ref{P2_c_1}) as follows
\begin{align}
&\mathbf{w}_{IE}^H \mathbf{B}_{8,g}\mathbf{w}_{IE}\label{w_MM}\\
&\geq
2\text{Re}\{ \mathbf{w}_{IE,0}^H \mathbf{B}_{8,g}(\mathbf{w}_{IE}-\mathbf{w}_{IE,0}) \}
+
\mathbf{w}_{IE,0}^H \mathbf{B}_{8,g}\mathbf{w}_{IE,0},\nonumber\\
&=2\text{Re}\{ \mathbf{b}_{4,g}^H\mathbf{w}_{IE} \}+c_{3,g},\nonumber
\end{align}
where $\mathbf{w}_{IE,0}$ is a feasible solution obtained in the last iteration,
$\mathbf{b}_{4,g} \triangleq \mathbf{B}_{8,g}^H\mathbf{w}_{IE,0}  $,
and $c_{3,g}$ is constant.
Therefore,
based on the above linearization process,
the term $\mathbf{w}_{IE}^H \mathbf{B}_{8,g}\mathbf{w}_{IE}$ can be replaced by (\ref{w_MM}),
and problem (P2) can be formulated as
\begin{subequations}
\begin{align}
\textrm{(P3)}:&\mathop{\textrm{max}}
\limits_{\mathbf{w}_{IE}
}\
-\mathbf{w}_{IE}^H\mathbf{B}_5\mathbf{w}_{IE}+2\text{Re}\{ \mathbf{b}_3^H\mathbf{w}_{IE} \} + c_2
\label{P3_obj}\\
\textrm{s.t.}\ 
& -2\text{Re}\{ \mathbf{b}_{4,g}^H\mathbf{w}_{IE} \} + \bar{c}_{3,g} \leq 0, \forall g \in \mathcal{G}, \label{P3_c_1}\\
&\mathbf{w}_{IE}^H\mathbf{w}_{IE}\leq P_t,\label{P3_c_2}
\end{align}
\end{subequations}
where
$\bar{c}_{3,g} \triangleq -{c}_{3,g}+Q_t $.
Problem (P3) is a typical second-order cone program (SOCP),
for which we can employ the standard numerical solvers, e.g., CVX \cite{ref_CVX}, to solve it.

However,
the dimensionality of the variable $\mathbf{w}_{IE}$, 
which scales with $L(K+G)$, can be very large.
As a result, the complexity of solving (P3) via CVX can be prohibitively high.
This observation motivates us to develop a more efficient optimization algorithm.

First,
by introducing $G+1$ auxiliary variables of the variable $\mathbf{w}_{IE}$,
i.e., $\mathbf{w}_{IE}=\mathbf{f}_{g}$, $\forall g \in \mathcal{\bar{G}} \triangleq 0 \cup \mathcal{G} $,
we decouple the $G+1$ constraints of problem (P3) as follows
\begin{subequations}
\begin{align}
\textrm{(P4)}:&\mathop{\textrm{min}}
\limits_{\mathbf{w}_{IE}, 
\{\mathbf{f}_{g}\}
}\
\mathbf{w}_{IE}^H\mathbf{B}_5\mathbf{w}_{IE}-2\text{Re}\{ \mathbf{b}_3^H\mathbf{w}_{IE} \} - c_2
\label{P4_obj}\\
\textrm{s.t.}\ 
& -2\text{Re}\{ \mathbf{b}_{4,g}^H\mathbf{f}_{g} \} + \bar{c}_{3,g} \leq 0, \forall g \in \mathcal{G}, 
\label{P4_c_1}\\
&\mathbf{f}_{0}^H\mathbf{f}_{0}\leq P_t,
\label{P4_c_2}\\
&\mathbf{w}_{IE} = \mathbf{f}_{g}, \forall g \in \mathcal{\bar{G}}.
\label{P4_c_3}
\end{align}
\end{subequations}

Next,
based on the ADMM methodology \cite{ref_ADMM},
we propose an iterative closed-form solution to update $\mathbf{w}_{IE}$.
By penalizing the equality constraints (\ref{P4_c_3}) into the objective (\ref{P4_obj}),
the augmented Lagrangian (AL) function of (P4) is given as
\begin{align}
&\mathcal{L}(\mathbf{w}_{IE},\mathbf{f}, \boldsymbol{\lambda})
=
\mathbf{w}_{IE}^H\mathbf{B}_5\mathbf{w}_{IE}-2\text{Re}\{ \mathbf{b}_3^H\mathbf{w}_{IE} \} - c_2\\
&+ \frac{\rho}{2}\bigg({\sum}_{g=0}^{G}\Vert\mathbf{w}_{IE} - \mathbf{f}_{g} \Vert_2^2\bigg)
+{\sum}_{g=0}^{G}\text{Re}\{ \boldsymbol{\lambda}_g^H (\mathbf{w}_{IE} - \mathbf{f}_{g} )\},\nonumber
\end{align}
where
$\mathbf{f} = [\mathbf{f}_0^T,\mathbf{f}_1^T,\cdots,\mathbf{f}_G^T]^T $,
$\boldsymbol{\lambda} = [\boldsymbol{\lambda}_0^T,\boldsymbol{\lambda}_1^T,\cdots,\boldsymbol{\lambda}_G^T]^T $
with each $\boldsymbol{\lambda}_g$ being introduced as the Lagrangian multiplier,
and $\rho$ is a positive constant.
Then (P4) can be rewritten as
\begin{subequations}
\begin{align}
\textrm{(P5)}:&\mathop{\textrm{min}}
\limits_{\mathbf{w}_{IE}, 
\mathbf{f},
\boldsymbol{\lambda}
}\
\mathcal{L}(\mathbf{w}_{IE},\mathbf{f}, \boldsymbol{\lambda})
\label{P5_obj}\\
\textrm{s.t.}\ 
& -2\text{Re}\{ \mathbf{b}_{4,g}^H\mathbf{f}_{g} \} + \bar{c}_{3,g} \leq 0, \forall g \in \mathcal{G}, 
\label{P5_c_1}\\
&\mathbf{f}_{0}^H\mathbf{f}_{0}\leq P_t.
\label{P5_c_2}
\end{align}
\end{subequations}

In the following,
within the ADMM framework,
we adopt the block coordinate descent (BCD) methodology to tackle (P5).
The update steps are presented below.

\underline{\textit{1) Update $\mathbf{w}_{IE}$:}}
When the variables $\mathbf{f}$ and $\boldsymbol{\lambda}$ are fixed, 
the subproblem with respect to (w.r.t.) $\mathbf{w}_{IE}$ can be written as
\begin{subequations}
\begin{align}
\textrm{(P6)}:&\mathop{\textrm{min}}
\limits_{\mathbf{w}_{IE}, 
}\
\mathbf{w}_{IE}^H \bar{\mathbf{B}}_{5} \mathbf{w}_{IE} - 2\text{Re}\{\bar{\mathbf{b}}_{3}^H\mathbf{w}_{IE}\}
\label{P6_obj}
\end{align}
\end{subequations}
where
$\bar{\mathbf{B}}_{5} \triangleq \mathbf{B}_{5} + \frac{\rho}{2}(G+1)\mathbf{I}_{L(K+G)}$,
{
$\bar{\mathbf{b}}_{3} \triangleq \mathbf{b}_{3} + \frac{\rho}{2}\big({\sum}_{g=0}^{G} \mathbf{f}_g \big)- \frac{1}{2} \big({\sum}_{g=0}^{G} \boldsymbol{\lambda}_g \big)$.}
We can observe that (P6) is an unconstrained convex quadratic problem.
Thus, its optimal solution can be directly obtained as
\begin{align}
\mathbf{w}_{IE}^{\star}=
\bar{\mathbf{B}}_{5}^{-1}\bar{\mathbf{b}}_{3}. \label{w_IE_closed_solution}
\end{align}

{It is worth noting that the matrix inversion involved in (\ref{w_IE_closed_solution}) is well-defined. 
The detailed analysis of the invertibility and numerical stability of 
$\bar{\mathbf{B}}_5$ is provided in Appendix~\ref{app:matrix_inversion}.}

\underline{\textit{2) Update $\{\mathbf{f}_{g}\}$:}}
With the other variables fixed, 
the optimization of the variables $\{\mathbf{f}_{g}\}$ reduces to solving the following problem
\begin{subequations}
\begin{align}
\textrm{(P7)}:\mathop{\textrm{min}}
\limits_{
\{\mathbf{f}_g\}
}\
&\frac{\rho}{2}{\sum}_{g=1}^{G}\Vert\mathbf{w}_{IE} - \mathbf{f}_{g} \Vert_2^2\label{P7_obj}\\
&+{\sum}_{g=1}^{G}\text{Re}\{ \boldsymbol{\lambda}_g^H (\mathbf{w}_{IE} - \mathbf{f}_{g} )\}\nonumber
\\
\textrm{s.t.}\ 
& -2\text{Re}\{ \mathbf{b}_{4,g}^H\mathbf{f}_{g} \} + \bar{c}_{3,g} \leq 0, \forall g \in \mathcal{G}. 
\label{P7_c_1}
\end{align}
\end{subequations}

Note that the subproblems w.r.t. $\mathbf{f}_{g}$ are independent of each other.
Therefore, 
(P7) can be decomposed into $G$ independent subproblems, 
where the $g$-th subproblem is formulated as
\begin{subequations}
\begin{align}
\textrm{(P7$_g$)}:&\mathop{\textrm{min}}
\limits_{
\{\mathbf{f}_g\}
}\
\Vert \mathbf{f}_{g} \Vert_2^2
-2\text{Re}\{ (\rho^{-1}\boldsymbol{\lambda}_g+\mathbf{w}_{IE})^H  \mathbf{f}_{g}\}
\label{P7_g_obj}\\
\textrm{s.t.}\ 
& -2\text{Re}\{ \mathbf{b}_{4,g}^H\mathbf{f}_{g} \} + \bar{c}_{3,g} \leq 0.
\label{P7_g_c_1}
\end{align}
\end{subequations}

Next,
(P7$_g$) can be solved by leveraging the following lemma, which is proved in \cite{ref_lemma_1}.
\begin{lemma}\label{lemma_1}
Consider the following optimization problem:
\begin{subequations}\label{P_Lem}
\begin{align}
\textrm{(P$_{\mathrm{Lem}}$)}:\ \min_{\mathbf{x}}\ & \mathbf{x}^H \mathbf{B} \mathbf{x}
-2\text{Re}\{\mathbf{b}^H\mathbf{x}\}-b \\
\mathrm{s.t.}\ & \mathbf{x}^H \bar{\mathbf{B}} \mathbf{x}
-2\text{Re}\{\bar{\mathbf{b}}^H\mathbf{x}\}-\bar{b}\le 0,
\end{align}
\end{subequations}
where $\mathbf{B}\succ\mathbf{0}$, $\bar{\mathbf{B}}\succeq\mathbf{0}$, and Slater's condition holds.
Then, the optimal solution to (P$_{\mathrm{Lem}}$) admits the following closed-form characterization:
\begin{itemize}
\item[] \underline{CASE-I}:
If we have
\begin{align}
(\mathbf{B}^{-1}\mathbf{b})^H\bar{\mathbf{B}}(\mathbf{B}^{-1}\mathbf{b})
\!-\!2\text{Re}\{\bar{\mathbf{b}}^H(\mathbf{B}^{-1}\mathbf{b})\}-\bar{b}\leq 0,
\end{align}
then the constraint is inactive at the optimum and
\begin{align}
\mathbf{x}^\star=\mathbf{B}^{-1}\mathbf{b}.
\end{align}

\item[] \underline{CASE-II}:
 Otherwise, the constraint is active and
\begin{align}
\mathbf{x}^\star=(\mu^\star\bar{\mathbf{B}}+\mathbf{B})^{-1}(\mu^\star\bar{\mathbf{b}}+\mathbf{b}),
\end{align}
where $\mu^\star>0$ is the unique value of
\begin{align}
&\big((\mu\bar{\mathbf{B}}+\mathbf{B})^{-1}(\mu\bar{\mathbf{b}}+\mathbf{b})\big)^H
\bar{\mathbf{B}}
(\mu^\star\bar{\mathbf{B}}+\mathbf{B})^{-1}(\mu^\star\bar{\mathbf{b}}+\mathbf{b})\nonumber\\
&\quad-2\text{Re}\{\bar{\mathbf{b}}^H(\mu^\star\bar{\mathbf{B}}+\mathbf{B})^{-1}
(\mu^\star\bar{\mathbf{b}}+\mathbf{b})\}-\bar{b}=0,
\end{align}
which can be efficiently obtained via Newton's method.
\end{itemize}
\end{lemma}

By using the above Lemma, 
we can efficiently update all the $G$ variables $\{\mathbf{f}_g\}$ in a parallel manner.

\underline{\textit{3) Update $\mathbf{f}_{0}$:}}
Next,
we will investigate the update of $\mathbf{f}_0$.
The subproblem w.r.t. $\mathbf{f}_0$ is given as
\begin{subequations}
\begin{align}
\textrm{(P8)}:&\mathop{\textrm{min}}
\limits_{
\mathbf{f}_0
}\
\Vert \mathbf{f}_{0} \Vert_2^2
-2\text{Re}\{ (\rho^{-1}\boldsymbol{\lambda}_0+\mathbf{w}_{IE})^H  \mathbf{f}_{0}\}
\label{P8_obj}\\
\textrm{s.t.}\ 
&\mathbf{f}_{0}^H\mathbf{f}_{0}\leq P_t.
\label{P8_c_1}
\end{align}
\end{subequations}

Next, 
a closed-form solution to (P8) can be derived by applying the Lagrangian multiplier method \cite{ref_Convex Optimization}. 
The details can be found in Appendix~\ref{app:P8_solution}.

\underline{\textit{4) Update $\boldsymbol{\lambda}$:}}
Following the ADMM procedure, 
after updating the primal variables,
the Lagrangian dual variables $\boldsymbol{\lambda}$ are updated as follows
\begin{align}
\boldsymbol{\lambda}_g^{(t+1)} := \boldsymbol{\lambda}_g^{(t)} + \rho(\mathbf{w}_{IE}-\mathbf{f}_g), 
\forall g \in \mathcal{\bar{G}}.\label{ADMM_rho}
\end{align}

Algorithm \ref{alg:1} summarizes the ADMM-based algorithm for solving (P3).
\begin{algorithm}[t]
\caption{The ADMM-Based Method to Solve (P3)}
\label{alg:1}
\begin{algorithmic}[1]
\STATE {initialize}
$\mathbf{w}_{IE}^{(0)}$,
$\mathbf{f}^{(0)}$,
$\boldsymbol{\lambda}^{(0)}$,
and
$t=0$;
\REPEAT
\STATE update $\mathbf{w}_{IE}^{(t)}$ by (\ref{w_IE_closed_solution});
\FOR{ $g = 1:G$ }
\STATE update $\mathbf{f}_{g}^{(t)}$ by solving (P7$_g$);
\ENDFOR
\STATE update $\mathbf{f}_{0}^{(t)}$ by solving (P8);
\FOR{ $g = 0:G$ }
\STATE {update $\boldsymbol{\lambda}_g^{(t)}$ by (\ref{ADMM_rho});}
\ENDFOR
\STATE $t++$;
\UNTIL{$convergence$;}
\end{algorithmic}
\end{algorithm}

\subsection{Updating The Holographic Beamformer}

In this subsection, 
we will discuss the optimization of the holographic beamforming $\boldsymbol{\psi}$.
First,
we rewrite the function  into a compact form of $\boldsymbol{\psi}$ as follows
\begin{align}
&\mathrm{\tilde{R}}_{k}
\Leftrightarrow
-\boldsymbol{\psi}^T( \mathbf{B}_{6,k}+\mathbf{B}_{7,k} )\boldsymbol{\psi}
+2\mathbf{b}_{5,k}^T\boldsymbol{\psi} + c_{4,k},
\nonumber
\end{align}
where the new coefficients are given in (\ref{mm_R_coefficient_1}).
\begin{figure*}
\begin{align}
&\mathbf{B}_{6,k} \triangleq \text{Re}\bigg\{ \omega_k\vert \beta_k \vert^2\bigg( 
{\sum}_{i=1}^{K}\text{diag}( \mathbf{Q}\mathbf{w}_{I,i} )^H \mathbf{C}_{M}^H\mathbf{h}_{I,k}
\mathbf{h}_{I,k}^H\mathbf{C}_{M}\text{diag}( \mathbf{Q}\mathbf{w}_{I,i} )
\bigg)    \bigg\}, \label{mm_R_coefficient_1} \\
&\mathbf{B}_{7,k} \triangleq \text{Re}\bigg\{ \omega_k\vert \beta_k \vert^2\bigg( 
{\sum}_{g=1}^{G}\text{diag}( \mathbf{Q}\mathbf{w}_{E,g} )^H \mathbf{C}_{M}^H\mathbf{h}_{I,k}
\mathbf{h}_{I,k}^H\mathbf{C}_{M}\text{diag}( \mathbf{Q}\mathbf{w}_{E,g} )
\bigg)    \bigg\} \nonumber,\\
&\mathbf{b}_{5,k} \triangleq
\text{Re}\{
\omega_{k}\beta_{k}^{\ast}
\text{diag}(\mathbf{Q}\mathbf{w}_{I,k})^T\mathbf{C}_{M}^T\mathbf{h}_{I,k}^{\ast}
\},
c_{4,k} \triangleq \log(\omega_k)-\omega_k+\omega_k\vert\beta_k \vert^2\sigma_{I,k}^2+1.
\nonumber
\end{align}
\boldsymbol{\hrule}
\end{figure*}
Therefore,
the objective (\ref{P1_obj}) of (P1) can be given as
\begin{align}
&{\sum}_{k=1}^{K} \tilde{\mathrm{R}}_{k}
\Leftrightarrow
-\boldsymbol{\psi}^T\mathbf{B}_{8}\boldsymbol{\psi}
+2\mathbf{b}_{6}^T\boldsymbol{\psi} + c_{5},
\nonumber
\end{align}
where
$\mathbf{B}_{8} \triangleq {\sum}_{k=1}^{K}(\mathbf{B}_{6,k}+\mathbf{B}_{7,k})$,
$\mathbf{b}_{6} \triangleq {\sum}_{k=1}^{K} \mathbf{b}_{5,k}$, 
$c_{5} \triangleq {\sum}_{k=1}^{K} c_{4,k}$.

Furthermore,
the constraints (\ref{P1_c_1}) can be reformulated as
\begin{align}
&\mathrm{Q}_g(\{\mathbf{w}_{I,k}\}, \{\mathbf{w}_{E,g}\}, \boldsymbol{\Psi})  \geq Q_t
\Leftrightarrow 
\boldsymbol{\psi}^T\mathbf{B}_{11,g}\boldsymbol{\psi}
\geq Q_t,
\end{align}
where
$\mathbf{B}_{11,g} \triangleq \mathbf{B}_{9,g} + \mathbf{B}_{10,g}$,
and 
$\mathbf{B}_{9,g}$ and $\mathbf{B}_{10,g}$
are defined in (\ref{mm_R_coefficient_2}).
\begin{figure*}
\begin{align}
&\mathbf{B}_{9,g} \triangleq \text{Re}\bigg\{ \bigg( 
{\sum}_{i=1}^{G}\text{diag}( \mathbf{Q}\mathbf{w}_{E,i} )^H \mathbf{C}_{M}^H\mathbf{h}_{E,g}
\mathbf{h}_{E,g}^H\mathbf{C}_{M}\text{diag}( \mathbf{Q}\mathbf{w}_{E,i} )
\bigg)    \bigg\}, \label{mm_R_coefficient_2} \\
&\mathbf{B}_{10,g} \triangleq \text{Re}\bigg\{ \bigg( 
{\sum}_{k=1}^{K}\text{diag}( \mathbf{Q}\mathbf{w}_{I,k} )^H \mathbf{C}_{M}^H\mathbf{h}_{E,g}
\mathbf{h}_{E,g}^H\mathbf{C}_{M}\text{diag}( \mathbf{Q}\mathbf{w}_{I,k} )
\bigg)    \bigg\}. \nonumber
\end{align}
\boldsymbol{\hrule}
\end{figure*}
Based on the above equivalent reformulation,
(P1) can be rewritten as
\begin{subequations}
\begin{align}
\textrm{(P9)}:&\mathop{\textrm{min}}
\limits_{ \boldsymbol{\psi}
}\
\boldsymbol{\psi}^T\mathbf{B}_{8} \boldsymbol{\psi}
-2\mathbf{b}_{6}^T\boldsymbol{\psi} - c_{5}
\label{P9_obj}\\
\textrm{s.t.}\ 
&\boldsymbol{\psi}^T\mathbf{B}_{11,g}\boldsymbol{\psi}  \geq Q_t, \forall g \in \mathcal{G}, \label{P9_c_1}\\
& 0 \leq [\boldsymbol{\psi}]_n \leq 1, \forall n \in \mathcal{N}. \label{P9_c_2}
\end{align}
\end{subequations}

Obviously, 
obtaining the optimal solution to (P9) is difficult
due to the non-convex constraints (\ref{P9_c_1}).
Inspired by the MM method,
the left-hand term of (\ref{P9_c_1}) can be convexified as follows
\begin{align}
&\boldsymbol{\psi}^T\mathbf{B}_{11,g}\boldsymbol{\psi}
\geq
2\boldsymbol{\psi}_0^T\mathbf{B}_{11,g}(\boldsymbol{\psi} - \boldsymbol{\psi}_0)
+\boldsymbol{\psi}_0^T\mathbf{B}_{11,g}\boldsymbol{\psi}_0 \label{psi_MM_1} \\
&=2\boldsymbol{\psi}_0^T\mathbf{B}_{11,g}\boldsymbol{\psi}
-\boldsymbol{\psi}_0^T\mathbf{B}_{11,g}\boldsymbol{\psi}_0.\nonumber
\end{align}

By replacing the left-hand term of (\ref{P9_c_1}) with (\ref{psi_MM_1}), 
problem (P9) is rewritten as
\begin{subequations}
\begin{align}
\textrm{(P10)}:&\mathop{\textrm{min}}
\limits_{ \boldsymbol{\psi}
}\
\boldsymbol{\psi}^T\mathbf{B}_{8} \boldsymbol{\psi}
-2\mathbf{b}_{6}^T\boldsymbol{\psi} - c_{5}
\label{P10_obj}\\
\textrm{s.t.}\ 
&-2\mathbf{b}_{7,g}^T\boldsymbol{\psi} + c_{6,g} \leq 0, \forall g \in \mathcal{G}, \label{P10_c_1}\\
& 0 \leq [\boldsymbol{\psi}]_n \leq 1, \forall n \in \mathcal{N}, \label{P10_c_2}
\end{align}
\end{subequations}
where 
$\mathbf{b}_{7,g} \triangleq \mathbf{B}_{11,g}^T\boldsymbol{\psi}_0 $
and $c_{6,g}$ is constant.
Problem (P10) is convex, and its optimal solution can be obtained by CVX.

Next, 
we propose a low-complexity algorithm to solve (P10) by leveraging ADMM. 
We first introduce the auxiliary variables 
$\boldsymbol{\phi} \triangleq [\boldsymbol{\phi}_0^T,\boldsymbol{\phi}_1^T,\cdots,\boldsymbol{\phi}_G^T]^T$ 
to decouple the constraints as follows
\begin{subequations}
\begin{align}
\textrm{(P11)}:&\mathop{\textrm{min}}
\limits_{ \boldsymbol{\psi}, \boldsymbol{\phi}
}\
\boldsymbol{\psi}^T\mathbf{B}_{8} \boldsymbol{\psi}
-2\mathbf{b}_{6}^T\boldsymbol{\psi} - c_{5}
\label{P11_obj}\\
\textrm{s.t.}\ 
&-2\mathbf{b}_{7,g}^T\boldsymbol{\phi}_g + c_{6,g} \leq 0, \forall g \in \mathcal{G}, 
\label{P11_c_1}\\
& 0 \leq [\boldsymbol{\phi}_0]_n \leq 1, \forall n \in \mathcal{N}, 
\label{P11_c_2}\\
& \boldsymbol{\psi} = \boldsymbol{\phi}_g, \forall g \in \bar{\mathcal{G}}.
\label{P11_c_2}
\end{align}
\end{subequations}

And the AL function of (P11) can be defined as
\begin{align}
&\mathcal{L}(\boldsymbol{\psi}, \boldsymbol{\phi}, \boldsymbol{\tau})
=
\boldsymbol{\psi}^T\mathbf{B}_{8} \boldsymbol{\psi} -2\mathbf{b}_{6}^T\boldsymbol{\psi} - c_{5}\\
&+ \frac{\eta}{2}\bigg({\sum}_{g=0}^{G}\Vert\boldsymbol{\psi} - \boldsymbol{\phi}_{g} \Vert_2^2\bigg)
+{\sum}_{g=0}^{G}\{ \boldsymbol{\tau}_g^T (\boldsymbol{\psi} - \boldsymbol{\phi}_{g} )\},\nonumber
\end{align}
where
$ \boldsymbol{\tau} \triangleq [ \boldsymbol{\tau}_0^T, \boldsymbol{\tau}_1^T, \cdots ,\boldsymbol{\tau}_G^T]^T$ are the Lagrangian multipliers
and 
$\eta$ is a constant term.
Thus,
we turn to solve the following problem
\begin{subequations}
\begin{align}
\textrm{(P12)}:&\mathop{\textrm{min}}
\limits_{ \boldsymbol{\psi}, \boldsymbol{\phi}, \boldsymbol{\tau}
}\
\mathcal{L}(\boldsymbol{\psi}, \boldsymbol{\phi}, \boldsymbol{\tau})
\label{P12_obj}\\
\textrm{s.t.}\ 
&-2\mathbf{b}_{7,g}^T\boldsymbol{\phi}_g + c_{6,g} \leq 0, \forall g \in \mathcal{G}, 
\label{P12_c_1}\\
& 0 \leq [\boldsymbol{\phi}_0]_n \leq 1, \forall n \in \mathcal{N}. 
\label{P12_c_2}
\end{align}
\end{subequations}

Following the optimization process of (P5), 
we can proceed to present the update steps in minimizing $\mathcal{L}(\boldsymbol{\psi}, \boldsymbol{\phi}, \boldsymbol{\tau})$ by using the BCD method.
And,
we will update variables $\boldsymbol{\psi}$, $\{\boldsymbol{\phi}_g\}$, 
$\boldsymbol{\phi}_0$,
and $\boldsymbol{\tau}$ sequentially.

\uwave{\textit{1) Update $\boldsymbol{\psi}$:}}
When other variables are fixed,
the update of the variable $\boldsymbol{\psi}$ can be done by solving the following unconstrained convex problem
\begin{subequations}
\begin{align}
\textrm{(P13)}:&\mathop{\textrm{min}}
\limits_{ \boldsymbol{\psi}
}\
\boldsymbol{\psi}^T\bar{\mathbf{B}}_{8} \boldsymbol{\psi}
-2\bar{\mathbf{b}}_{6}^T\boldsymbol{\psi}
\label{P13_obj}
\end{align}
\end{subequations}
where
$\bar{\mathbf{B}}_{8} \triangleq {\mathbf{B}}_{8} + \frac{\eta}{2}(G+1)\mathbf{I}_N$,
$\bar{\mathbf{b}}_{6} \triangleq {\mathbf{b}}_{6} + \frac{\eta}{2}\big({\sum}_{g=0}^{G}\boldsymbol{\phi}_g \big) 
- \frac{1}{2}\big({\sum}_{g=0}^{G}\boldsymbol{\tau}_g \big)$.
Hence,
we can directly obtain the optimal solution of $\boldsymbol{\psi}$ as follows
\begin{align}
\boldsymbol{\psi}^{\star}=\bar{\mathbf{B}}_{8}^{-1}\bar{\mathbf{b}}_{6}.\label{psi_closed_solution}
\end{align}

{
Since $\bar{\mathbf{B}}_8=\mathbf{B}_8+\frac{\eta}{2}(G+1)\mathbf{I}_{N}$ with $\mathbf{B}_8\succeq \mathbf{0}$ and $\eta>0$, 
$\bar{\mathbf{B}}_8$ is positive definite and nonsingular following the same argument as in Appendix~\ref{app:matrix_inversion}.
}

\uwave{\textit{2) Update $\{\boldsymbol{\phi}_g\}$:}}
To update $\{\boldsymbol{\phi}_g\}$, 
we need to solve the following problem
\begin{subequations}
\begin{align}
\textrm{(P14)}:\mathop{\textrm{min}}
\limits_{ \{\boldsymbol{\phi}_g\}
}\
&\frac{\eta}{2}{\sum}_{g=1}^{G}\!\Vert\boldsymbol{\psi}\!\! -\!\! \boldsymbol{\phi}_{g} \Vert_2^2
\!+\!\!{\sum}_{g=1}^{G}\!\{ \boldsymbol{\tau}_g^T (\boldsymbol{\psi}\!\! -\!\! \boldsymbol{\phi}_{g} )\}\label{P14_obj}\\
\textrm{s.t.}\ 
&-2\mathbf{b}_{7,g}^T\boldsymbol{\phi}_g + c_{6,g} \leq 0, \forall g \in \mathcal{G}. 
\label{P14_c_1}
\end{align}
\end{subequations}

Thus, the variables 
$\{\boldsymbol{\phi}_g\}$ 
in both the objective and the constraints are decoupled, 
and each variable $\boldsymbol{\phi}_g$ can be updated independently by solving the following problem
\begin{subequations}
\begin{align}
\textrm{(P14$_g$)}:&\mathop{\textrm{min}}
\limits_{ \{\boldsymbol{\phi}_g\}
}\
\Vert \boldsymbol{\phi}_{g} \Vert_2^2
-2 (\eta^{-1}\boldsymbol{\tau}_g + \boldsymbol{\psi} )^T  \boldsymbol{\phi}_{g} \\
\textrm{s.t.}\ 
&-2\mathbf{b}_{7,g}^T\boldsymbol{\phi}_g + c_{6,g} \leq 0. 
\end{align}
\end{subequations}

Obviously,
the above problem can be analytically solved via invoking Lemma \ref{lemma_1}.

\uwave{\textit{3) Update $\boldsymbol{\phi}_0$:}}
With other variables fixed, 
the subproblem w.r.t. $\boldsymbol{\phi}_0$ is
\begin{subequations}
\begin{align}
\textrm{(P15)}:&\mathop{\textrm{min}}
\limits_{ \boldsymbol{\phi}_0
}\
\Vert \boldsymbol{\phi}_{0} \Vert_2^2
-2 \mathbf{b}_{7}^T  \boldsymbol{\phi}_{0} 
\label{P15_obj}\\
\textrm{s.t.}\ 
&0 \leq [\boldsymbol{\phi}_0]_n \leq 1, \forall n \in \mathcal{N},
\label{P15_c_1}
\end{align}
\end{subequations}
where
$\mathbf{b}_{7} \triangleq \eta^{-1}\boldsymbol{\tau}_0 + \boldsymbol{\psi}$.
Furthermore,
(P15) can be split into $N$ element-wise subproblems,
which can be given as
\begin{subequations}
\begin{align}
\textrm{(P15$_n$)}:&\mathop{\textrm{min}}
\limits_{ [\boldsymbol{\phi}_0]_n
}\
([\boldsymbol{\phi}_0]_n)^2
-2 [\mathbf{b}_{7}]_n  [\boldsymbol{\phi}_0]_n
\label{P15_obj}\\
\textrm{s.t.}\ 
&0 \leq [\boldsymbol{\phi}_0]_n \leq 1.
\label{P15_c_1}
\end{align}
\end{subequations}

Note that problem (P15$_n$) is a convex quadratic program with simple bound constraints.
Therefore,
we can directly obtain the optimal value of $[\boldsymbol{\phi}_0]_n$ as follows
\begin{equation}
[\boldsymbol{\phi}_0^{\star}]_n=\label{phi_0_closed_solution}
\begin{cases}
0, & \text{if }  [\mathbf{b}_{7}]_n\leq 0,\\[4pt]
1, & \text{if } [\mathbf{b}_{7}]_n\geq 1,\\[4pt]
[\mathbf{b}_{7}]_n, &\text{otherwise}.
\end{cases}
\end{equation}

\uwave{\textit{4) Update $\boldsymbol{\tau}$:}}
The Lagrangian dual variables $\boldsymbol{\tau}$ can be updated as follows
\begin{align}
\boldsymbol{\tau}_g^{t+1} := \boldsymbol{\tau}_g^{t} 
+ \eta(\boldsymbol{\psi}-\boldsymbol{\phi}_g),
\forall g \in \mathcal{\bar{G}}.\label{ADMM_eta}
\end{align}

The ADMM-based method to efficiently update $\boldsymbol{\psi}$ is summarized in Algorithm \ref{alg:2}.
\begin{algorithm}[t]
\caption{The ADMM-Based Method to Solve (P11)}
\label{alg:2}
\begin{algorithmic}[1]
\STATE {initialize}
$\boldsymbol{\psi}^{(0)}$,
$\boldsymbol{\phi}^{(0)}$,
$\boldsymbol{\tau}^{(0)}$,
and
$t=0$;
\REPEAT
\STATE update $\boldsymbol{\psi}^{(t)}$ by (\ref{psi_closed_solution});
\FOR{ $g = 1:G$ }
\STATE update $\boldsymbol{\phi}_{g}^{(t)}$ by solving (P14$_g$);
\ENDFOR
\FOR{ $n = 1:N$ }
\STATE update $[\boldsymbol{\phi}_{0}^{(t)}]_n$ by (\ref{phi_0_closed_solution});
\ENDFOR
\FOR{ $g = 0:G$ }
\STATE update $\boldsymbol{\tau}_g^{(t)}$ by (\ref{ADMM_eta});
\ENDFOR
\STATE $t++$;
\UNTIL{$convergence$;}
\end{algorithmic}
\end{algorithm}
The overall procedure to solve (P1) is outlined in Algorithm \ref{alg:3}.
\begin{algorithm}[t]
\caption{Solving the Problem (P1)}
\label{alg:3}
\begin{algorithmic}[1]
\STATE {initialize}
$\mathbf{w}_{IE}^{(0)}$,
$\boldsymbol{\psi}^{(0)}$,
and
$t=0$;
\REPEAT
\STATE update $\{\beta_k^{(t+1)}\}$ and $\{\omega_k^{(t+1)}\}$ by (\ref{beta_opt}) and (\ref{omega_opt}), respectively;
\STATE update $\mathbf{w}_{IE}^{(t+1)}$     by solving  (P3);
\STATE {update $\boldsymbol{\psi}^{(t+1)}$   by solving  (P10);}
\STATE $t++$;
\UNTIL{$convergence$;}
\end{algorithmic}
\end{algorithm}

\subsection{Convergence, Feasible Initialization, and Complexity Analysis}
\textit{1) Convergence analysis:}
{
Alg.~\ref{alg:3} is guaranteed to generate a non-decreasing objective sequence for (P1). 
Specifically, for fixed beamforming variables, the auxiliary variables $\{\beta_k\}$ and $\{\omega_k\}$ are updated optimally according to (\ref{beta_opt}) and (\ref{omega_opt}). 
For the beamforming updates, 
the MM-based surrogate constraints in (\ref{w_MM}) and (\ref{psi_MM_1}) are tight at the current iterate, 
and hence the previous iterate remains feasible to the corresponding surrogate subproblems (P3) and (P10).
Therefore, solving these subproblems exactly cannot decrease the objective value. 
Moreover, due to the transmit power constraint, the amplitude constraint, and the EH constraints, the feasible set of $(\mathbf{w}_{IE},\boldsymbol{\psi})$ is compact, 
which implies that the objective sequence is upper-bounded and thus convergent. 
By standard MM/BCA arguments, any limit point of $\{\mathbf{w}_{IE}^{(t)}, \boldsymbol{\psi}^{(t)}\}$ is a stationary point of (P1).
}

\textit{2) Feasible initialization:}
{
Since the MM-based surrogate constraints are constructed around the current
iterate, Alg.~\ref{alg:3} requires a feasible initial point. To obtain such a point,
we first consider the following feasibility problem
\begin{subequations}
\begin{align}
\textrm{(P$_{init,1}$)}:&\mathop{\textrm{Find}}\
(\{\mathbf{w}_{I,k}\}, \{\mathbf{w}_{E,g}\}, \boldsymbol{\Psi})\\
\textrm{s.t.}\ 
&\mathrm{Q}_g(\{\mathbf{w}_{I,k}\}, \{\mathbf{w}_{E,g}\}, \boldsymbol{\Psi})  \geq Q_t, \forall g \in \mathcal{G}, \label{P_init_1_c_1}\\
&{\sum}_{k=1}^{K}\mathbf{w}_{I,k}^H\mathbf{w}_{I,k}
+{\sum}_{g=1}^{G}\mathbf{w}_{E,g}^H\mathbf{w}_{E,g}\leq P_t,\label{P_init_1_c_2}\\
& 0 \leq [\boldsymbol{\psi}]_n \leq 1, \forall n \in \mathcal{N}. \label{P_init_1_c_3}
\end{align}
\end{subequations}

The above problem is intractable due to its missing objective. 
Therefore, we equivalently consider the following problem:
\begin{subequations}
\begin{align}
\textrm{(P$_{init,2}$)}:&\mathop{\textrm{max}}
\limits_{ \{\mathbf{w}_{I,k}\}, \{\mathbf{w}_{E,g}\}, \boldsymbol{\Psi},\xi
}\
\xi\\
\textrm{s.t.}\ 
&\mathrm{Q}_g(\{\mathbf{w}_{I,k}\}, \{\mathbf{w}_{E,g}\}, \boldsymbol{\Psi})  - Q_t \geq \xi
, \forall g \in \mathcal{G}, \\
& (\text{\ref{P_init_1_c_2}})-(\text{\ref{P_init_1_c_3}}).
\end{align}
\end{subequations}

It is readily observed that problem (P$_{ init,1}$) has a feasible solution
if and only if the optimal value of problem (P$_{ init,2}$) is nonnegative.
Specifically, if $\xi^\star \geq 0$, then the corresponding
$(\{\mathbf{w}_{I,k}^{\star}\}, \{\mathbf{w}_{E,g}^{\star}\},
\boldsymbol{\Psi}^{\star})$ satisfies all constraints of
(P$_{ init,1}$) and can be used as the initial point for Alg.~\ref{alg:3}.}

\textit{3) Complexity analysis:}
{
In this subsection, 
we analyze the computational complexity of the proposed algorithms. 
We first consider the SOCP-based method. 
According to \cite{ref_Complexity},
the dominant complexity of solving problem (P3) is given by
$\mathcal{O}((L(K+G))^{3.5})$.
Similarly, 
the dominant complexity of solving problem (P10) is given as
$\mathcal{O}(N^{3.5})$.
Therefore, 
the overall complexity of the SOCP-based method can be approximately expressed as
$
\mathcal{O}(
I_{\rm out}
(
(L(K+G))^{3.5}+N^{3.5}
)
)$,
where $I_{\rm out}$ denotes the number of outer iterations.

Next, we analyze the complexity of the proposed low-complexity ADMM-based method. 
For solving problem (P3), 
the complexity  is approximately given by
$\mathcal{O}(
L^3(K+G)^3
)$.
The dominant complexity for solving problem (P10) is given as 
$\mathcal{O}(N^3)$.
}

\section{Numerical Results}

\begin{figure}[t]
	\centering
	\includegraphics[width=.30\textwidth]{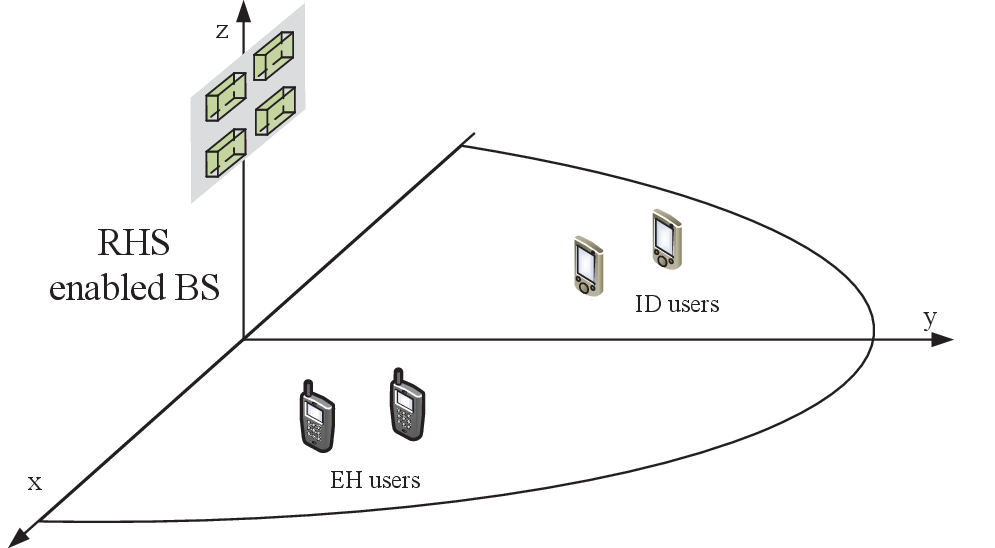}
	\caption{Simulation setup of the considered RHS-enabled SWIPT system.}
	\label{fig.3_1}
\end{figure}

{
In this section, we perform simulations to evaluate the performance of the proposed RHS-enabled SWIPT system.
As shown in Fig.~\ref{fig.3_1}, the considered system consists of one BS, $K=2$ ID users, and $G=2$ EH users.
The BS is equipped with $L=6$ RF chains connected to an RHS with $N=5\times 5$ radiation elements.
The BS is located at the three-dimensional coordinate $(0,0,4.5)$ m.
Both ID and EH users are placed at a height of $1.5$ m and are independently and uniformly distributed within a sector.
For the adopted linear EH model, the receiver noise at the EH users is neglected in the harvested-energy calculation.
Unless otherwise specified, the remaining simulation parameters are summarized in Table~\ref{tab:sim_parameters}.
}

\begin{table}[t]
\centering
\caption{{Simulation parameters.}}
\label{tab:sim_parameters}
\begin{tabular}{|c|c|}
\hline
\textbf{Parameter} & \textbf{Value} \\ \hline
Carrier frequency, $f_c$ & $30~\mathrm{GHz}$ \\ \hline
Wavelength, $\lambda$ & $0.01~\mathrm{m}$ \\ \hline
Inter-element spacing, $d$ & $\lambda/4 = 0.0025~\mathrm{m}$ \\ \hline
Refractive index, $\kappa$ & $\sqrt{3}$ \\ \hline
Equivalent RHS unit length, $l_d$ & $\lambda/10 = 0.001~\mathrm{m}$ \\ \hline
Sector angle, $\Theta$ & $180^\circ$ \\ \hline
ID-user distance range & $20-50~\mathrm{m}$ \\ \hline
Path-loss exponent for ID links, $\alpha_I$ & 3.2 \\ \hline
Path-loss exponent for EH links, $\alpha_E$ & 2.2 \\ \hline
Channel model & Rician fading \\ \hline
Rician factor, $K_R$ & $5~\mathrm{dB}$ \\ \hline
Maximum BS transmit power, $P_t$ & $10~\mathrm{dBm}$ \\ \hline
Noise power at ID users, $\sigma_{I,k}^2$ & $-80~\mathrm{dBm}$ \\ \hline
Energy harvesting efficiency, $\zeta$ & 1 \\ \hline
Minimum EH threshold, $Q_t$ & $0.1~\mu W$ \\ \hline
Source impedance, $Z_O$ & $50~\Omega$ \\ \hline
Antenna impedance, $Z_{A,T}$ & $0.8-j17.8~\Omega$ \\ \hline
\end{tabular}
\end{table}

\begin{figure}[t]
	\centering
	\includegraphics[width=.37\textwidth]{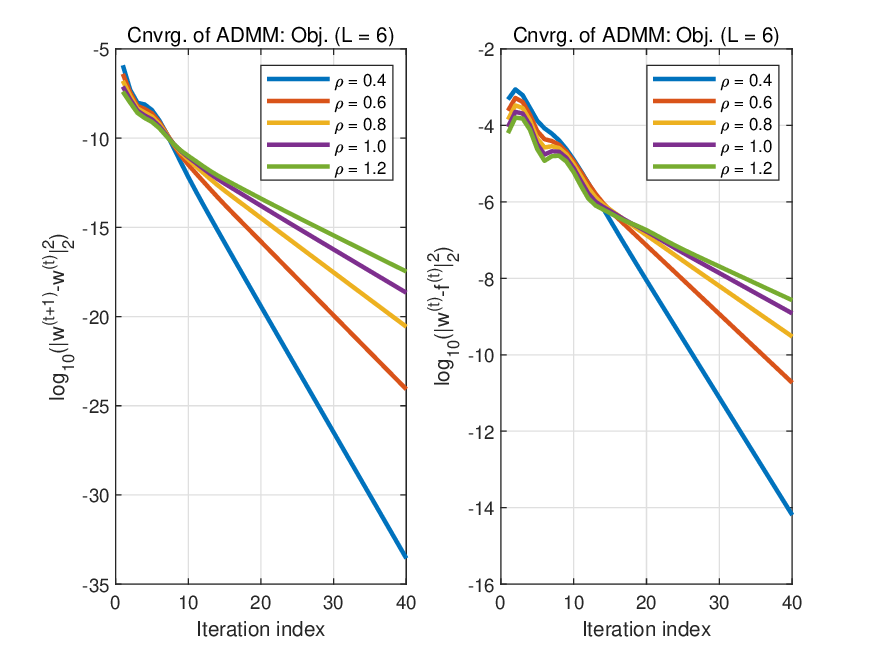}
	\caption{Convergence of the ADMM algorithm to solve (P3).}
	\label{fig.3}
\end{figure}

\begin{figure}[t]
	\centering
	\includegraphics[width=.37\textwidth]{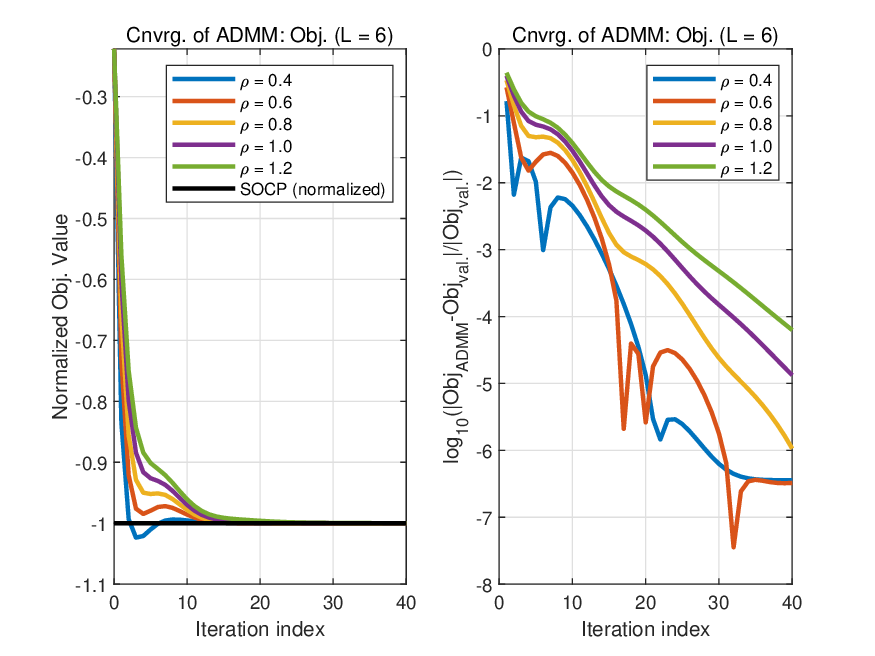}
	\caption{Convergence of the objective value of the ADMM method to solve (P3).}
	\label{fig.4}
\end{figure}

Both Figs. \ref{fig.3} and \ref{fig.4} illustrate the convergence behavior of 
our proposed ADMM-based method (i.e., Alg. \ref{alg:1}) for updating the digital beamforming.
In Fig. \ref{fig.3},
under various settings of constant $\rho$,
the left and right subfigures show  the differences
$\vert \mathbf{w}^{(t+1)}-\mathbf{w}^{(t)} \vert_2^2$
and 
$\vert\mathbf{w}^{(t)}-\mathbf{f}^{(t)} \vert_2^2$
in the logarithmic domain, respectively, along with the progress of ADMM iterations.
As shown in Fig. \ref{fig.3},
we observe that an appropriate value of $\rho$, 
which lies in the range of [0.6, 1.4], 
can achieve high-quality convergence 
(e.g., to a precision of $10^{-8}$) within $40$ iterations.
These curves illustrate the impact of parameter $\rho$ on the convergence rate and the stability of the ADMM-based algorithm.

The objective convergence behavior yielded by Alg.~\ref{alg:1} is shown in Fig. \ref{fig.4}.
The left subfigure of Fig.~\ref{fig.4} presents the evolution of the objective value under different settings of 
the coefficient $\rho$.
The value of the black line can be obtained by solving (P4) using CVX.
After normalization, it can be used as a benchmark for comparison.
The right subfigure illustrates the difference, in the logarithmic domain, 
between the objective values produced by CVX and Alg.~\ref{alg:1}.
As can be observed from Fig.~\ref{fig.4}, 
Alg.~\ref{alg:1} can yield sufficiently accurate objective values within $40$ iterations 
and attain a $10^{-4}$ precision.

\begin{figure}[t]
	\centering
	\includegraphics[width=.37\textwidth]{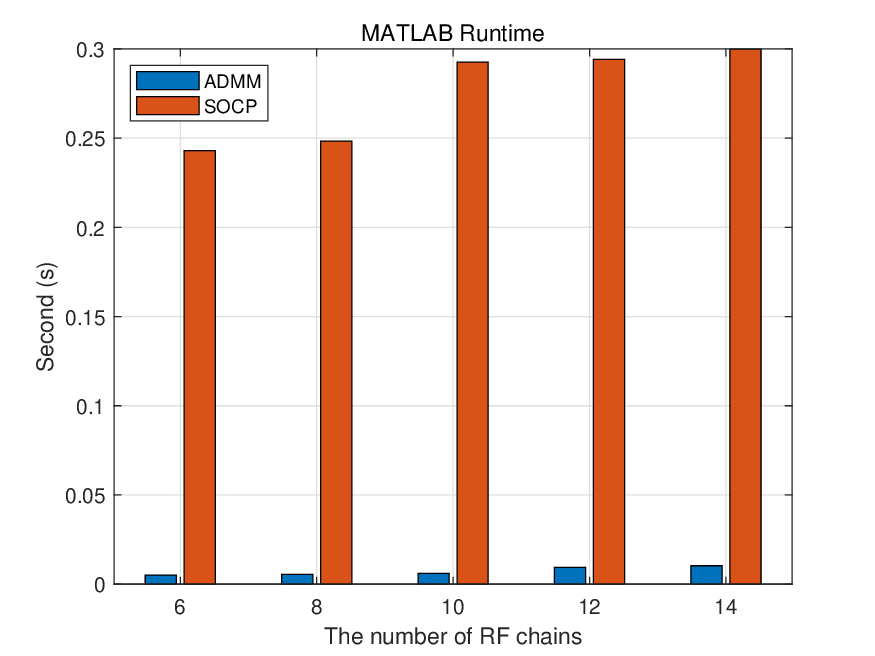}
	\caption{MATLAB Runtime to Solve (P3) (in Sec.).}
	\label{fig.4_1}
\end{figure}

To examine the computational complexity of CVX and Alg. \ref{alg:1}, 
Fig. \ref{fig.4_1} reports the MATLAB runtime of both methods for different numbers of RF chains $L$.
First, it can be seen that the runtime of both CVX and Alg. \ref{alg:1} increases as the number of RF chains increases.
Furthermore, Alg. \ref{alg:1} consumes roughly one to two orders of magnitude less time than CVX.

\begin{figure}[t]
	\centering
	\includegraphics[width=.37\textwidth]{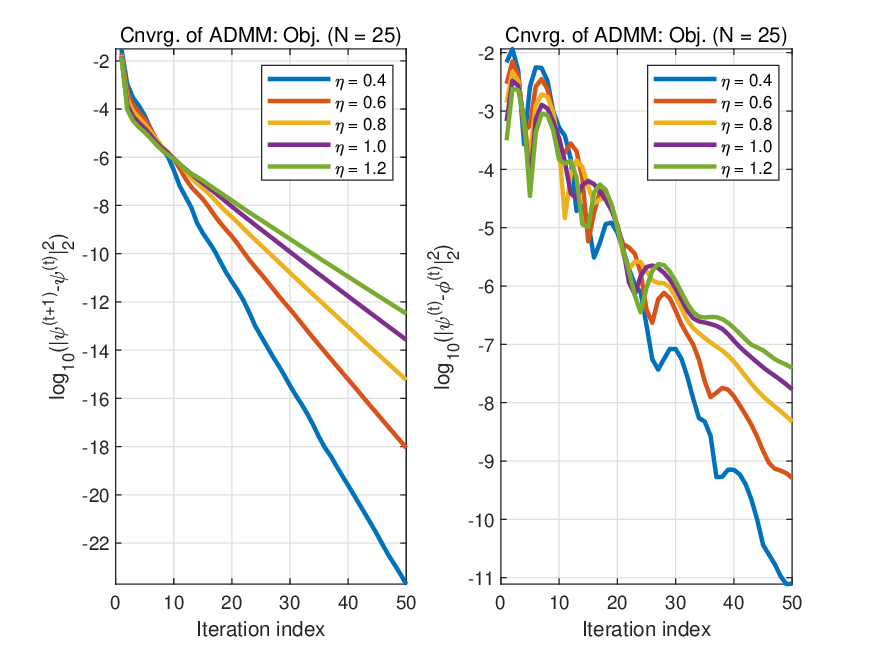}
	\caption{Convergence of the ADMM algorithm to solve (P10).}
	\label{fig.5}
\end{figure}

\begin{figure}[t]
	\centering
	\includegraphics[width=.37\textwidth]{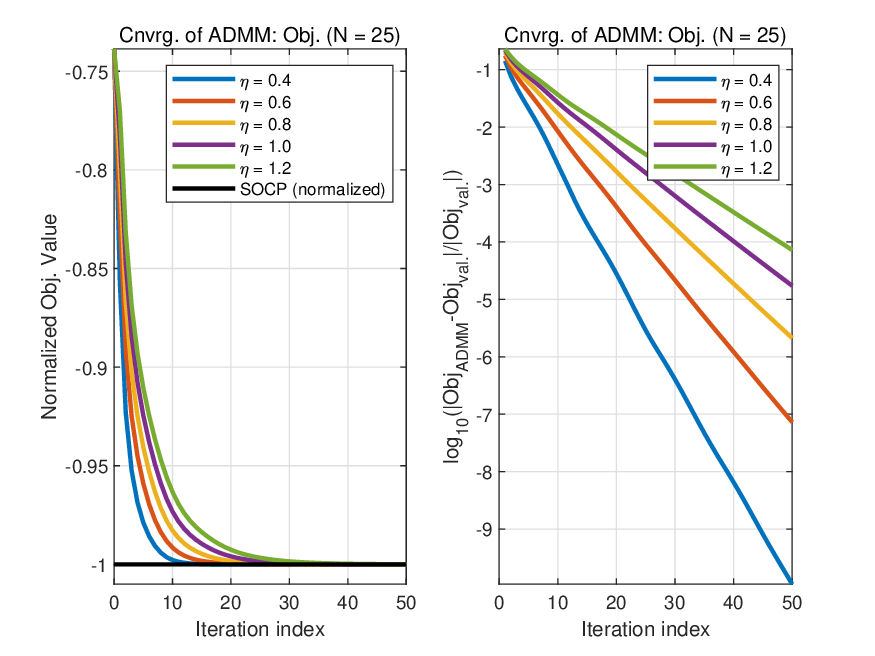}
	\caption{Convergence of the objective value of the ADMM method to solve (P10).}
	\label{fig.6}
\end{figure}

In Figs.~\ref{fig.5} and \ref{fig.6}, 
we examine the convergence behavior of the proposed ADMM-based method, 
which is invoked to optimize the holographic beamforming $\boldsymbol{\psi}$.
In Fig.~\ref{fig.5}, 
the differences
$\Vert\boldsymbol{\psi}^{(t+1)}-\boldsymbol{\psi}^{(t)}\Vert_2^2$ 
and $\Vert\boldsymbol{\psi}^{(t)}-\boldsymbol{\phi}^{(t)}\Vert_2^2$ 
are plotted in the left and right subfigures, respectively, under different values of $\eta$. 
With a suitable choice of $\eta$, 
the algorithm reaches a stationary point with respect to $\boldsymbol{\psi}$ within $50$ ADMM iterations, 
achieving a tolerance of $10^{-7}$. 
Moreover, when $\eta \in [0.4, 1.2]$, a smaller $\eta$ leads to faster convergence.

Fig.~\ref{fig.6} shows the convergence behavior of the objective value produced by Alg.~\ref{alg:2} under various ADMM parameter settings. 
Note that the black line in the left subfigure serves as the benchmark, 
which is obtained by solving (P10) using CVX. 
The right subfigure presents the difference between the objective values obtained by CVX and by Alg.~\ref{alg:2}. 
From the right part of Fig.~\ref{fig.6}, 
we observe that this difference drops below $10^{-4}$ within $50$ iterations,
which indicates that Alg.~\ref{alg:2} achieves performance comparable to CVX.

\begin{figure}[t]
	\centering
	\includegraphics[width=.37\textwidth]{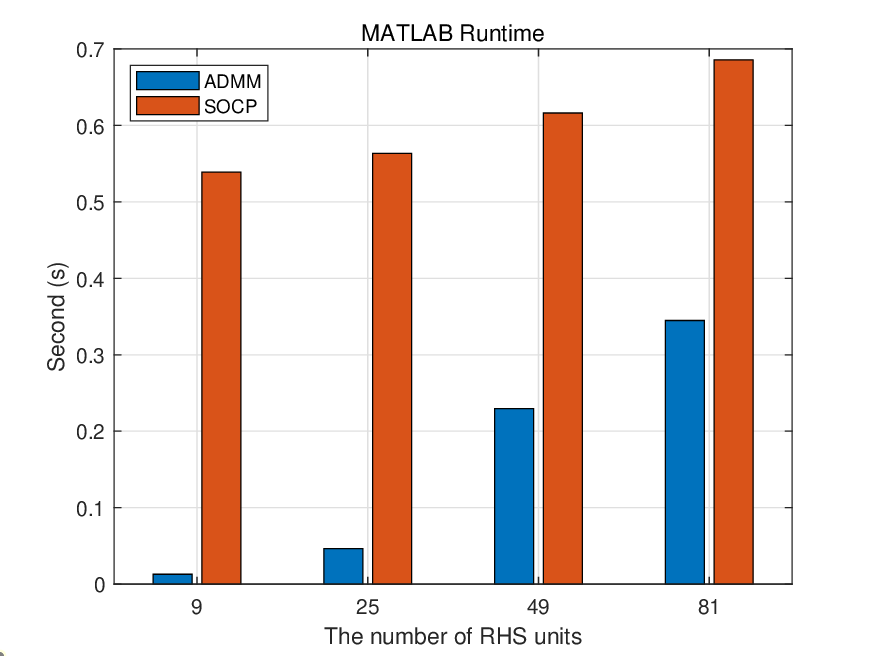}
	\caption{MATLAB Runtime to Solve (P10) (in Sec.).}
	\label{fig.6_1}
\end{figure}

Fig.~\ref{fig.6_1} lists the MATLAB runtime per iteration of CVX and Alg.~\ref{alg:2} 
for different numbers of RHS units. 
We observe that the runtime of both CVX and Alg.~\ref{alg:2} increases as the number of RHS units grows. 
Moreover, Alg.~\ref{alg:2} typically requires one to two orders of magnitude less runtime than CVX. 
Taken together with Figs.~\ref{fig.5} $-$ \ref{fig.6_1}, 
these results indicate that the proposed ADMM-based method (i.e., Alg.~\ref{alg:2}) 
achieves comparable performance while substantially reducing the computational complexity of solving (P10) compared with CVX.

\begin{figure}[t]
	\centering
	\includegraphics[width=.37\textwidth]{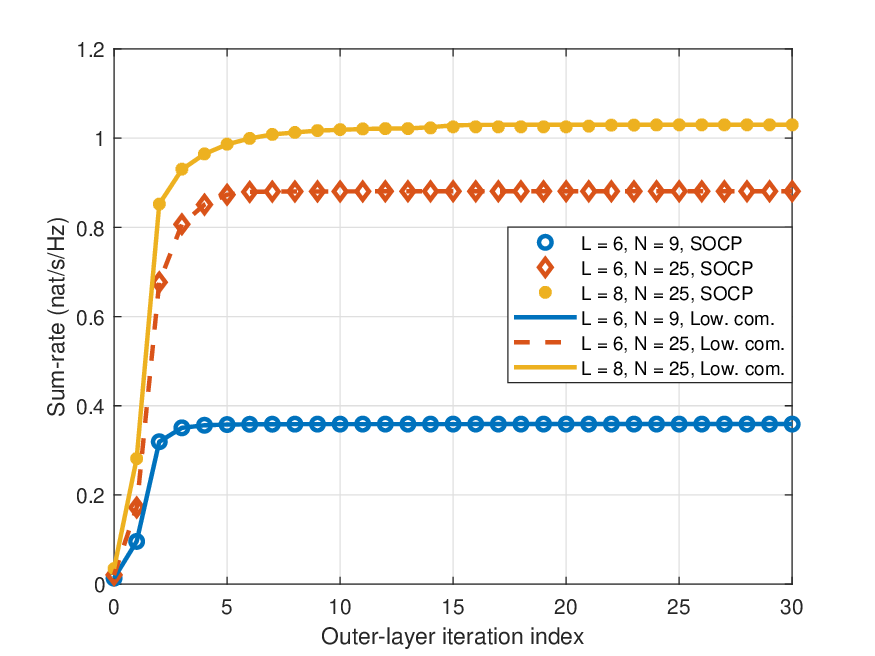}
	\caption{Convergence of both SOCP-based and low-complexity algorithms.}
	\label{fig.7}
\end{figure}

First, we label the SOCP-based and low-complexity methods as ``SOCP'' and ``Low. com.'', respectively. 
Fig.~\ref{fig.7} demonstrates the overall convergence of the proposed algorithms for solving (P1) 
under different system settings.
We observe that the sum-rate for both ``SOCP'' and ``Low. com.'' increases rapidly and converges within $20$ iterations. 
Moreover, 
the two methods achieve nearly identical performance across different system settings. 
Furthermore, the sum-rate increases as $N$ and/or $L$ increases.

\begin{figure}[t]
	\centering
	\includegraphics[width=.37\textwidth]{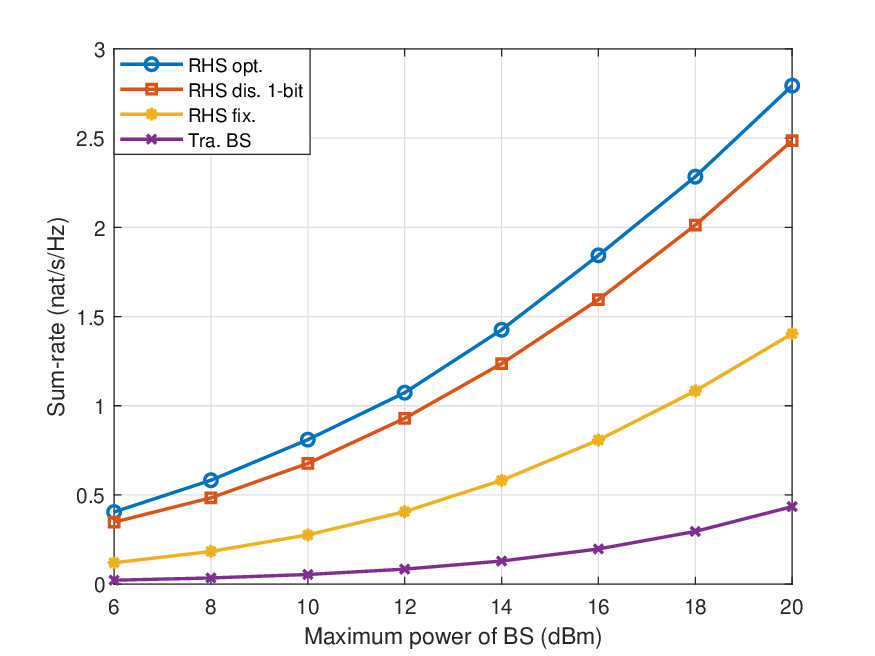}
	\caption{Sum-rate versus the maximum power.}
	\label{fig.8}
\end{figure}

{

First, 
to illustrate the efficiency of the proposed algorithm and the advantage of the RHS architecture,
we consider the following schemes:

1) 
\textbf{RHS opt.}: The BS digital beamformer and the RHS holographic beamformer are jointly optimized by the proposed Alg.~\ref{alg:3}. 
The transmit signal model follows (\ref{signal_model_1}) and (\ref{signal_model_2}), 
and the design is subject to the EH constraints in (\ref{P0_c_1}), 
the BS transmit power constraint in (\ref{P0_c_2}), 
and the RHS amplitude constraint in (\ref{P0_c_3}).

2) 
\textbf{RHS dis. 1-bit}: The system architecture and signal model are the same as those of ``RHS opt.'', while the RHS holographic beamformer is constrained to have finite-resolution amplitudes. 
Specifically, each diagonal entry of $\mathbf{\Psi}$ is restricted to the discrete set $\mathcal{A}_B=\{0,\frac{1}{2^B-1},\frac{2}{2^B-1},\ldots,1\}$, where $B$ denotes the amplitude resolution in bits. 
Under this discrete-amplitude constraint, 
the BS digital beamformers and the RHS holographic beamformer are jointly optimized, 
subject to the same EH constraints in (\ref{P0_c_1}) and BS transmit-power constraint in (\ref{P0_c_2}) as in ``RHS opt.''.

3) 
\textbf{RHS fix.}: The system architecture and signal model are the same as those of the proposed RHS-assisted system.
However, the RHS holographic beamformer is fixed as $\mathbf{\Psi}=\mathbf{\Psi}_{\rm fix}$ and is not optimized, 
where $\mathbf{\Psi}_{\rm fix}$ is set to the initial value of the RHS holographic beamformer used in the proposed ``RHS opt.'' scheme. 
Only the BS digital beamformers $\{\mathbf{w}_{I,k}\}$ and $\{\mathbf{w}_{E,g}\}$ are optimized under the same EH constraints and BS transmit-power constraint as in the proposed scheme. 
Since $\mathbf{\Psi}_{\rm fix}$ is taken from the feasible initialization of ``RHS opt.'', 
it also satisfies $0 \le [\boldsymbol{\psi}]_n \le 1$, $\forall n$.

4) 
\textbf{Tra.\ BS}: A conventional phased-array BS without RHS is considered. 
The transmitted signal is given as
$\mathbf{x}_{\rm Tra}=\sum_{k=1}^{K}\mathbf{v}_{I,k}s_{I,k}+\sum_{g=1}^{G}\mathbf{v}_{E,g}s_{E,g}$,
where $\mathbf{v}_{I,k}$ and $\mathbf{v}_{E,g}$ denote the digital beamformers for information and energy transmission, respectively. 
The conventional BS is assumed to have the same aperture as the RHS. 
Moreover, since a conventional phased-array structure with half-wavelength antenna spacing is adopted, 
the mutual coupling effect is ignored for this benchmark. 
Accordingly, the conventional BS is subject to the same total transmit power budget and EH requirements as the RHS-assisted schemes, while no RHS-related amplitude constraint is involved.
}

{
Fig.~\ref{fig.8} depicts the achieved sum-rate of four schemes versus the total transmit power $P_t$. 
Specifically, $P_t$ is varied from $6$ to $20$ dBm while keeping the other system parameters unchanged. 
As shown in Fig.~\ref{fig.8}, 
the achieved sum-rate of all schemes monotonically increases with $P_t$, 
since a larger transmit power generally enhances the effective received signal strength and thus improves the achievable rates. 
Among the four schemes, 
``RHS opt.'' consistently achieves the highest sum-rate over the entire power range, 
because it jointly optimizes the BS digital beamformer and the RHS holographic beamformer, 
thereby fully exploiting the spatial DoFs provided by the RHS. 
The scheme ``RHS dis. 1-bit'' attains a lower sum-rate than ``RHS opt.'', 
which reveals the performance loss caused by the finite-resolution amplitude constraint on the RHS holographic beamformer. 
Nevertheless, ``RHS dis. 1-bit'' still significantly outperforms ``RHS fix.'', 
demonstrating that even a low-resolution discrete-amplitude design can effectively improve the system performance. 
By contrast, ``RHS fix.'' suffers from a clear performance degradation since the RHS holographic beamformer is fixed 
and only the BS digital beamformer is optimized. 
Moreover, compared with ``Tra.~BS'', 
both ``RHS opt.'' and ``RHS dis. 1-bit'' achieve substantially higher sum-rates, 
which verifies the advantage of the RHS within the considered transmit power range.
}

\begin{figure}[t]
	\centering
	\includegraphics[width=.37\textwidth]{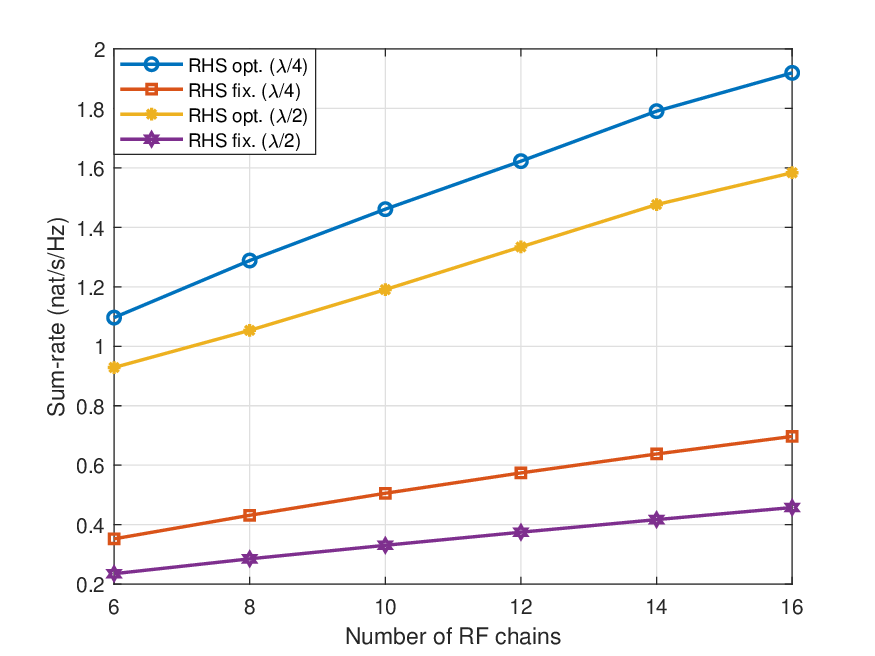}
	\caption{Sum-rate versus the number of RF chains.}
	\label{fig.9}
\end{figure}

To further evaluate the advantage of the holographic architecture with shorter antenna spacing, 
we consider two spacing configurations, 
i.e., $\frac{1}{2}\lambda$ and $\frac{1}{4}\lambda$. 
Fig.~\ref{fig.9} plots the achieved sum-rate versus the number of RF chains, $L$, 
under these two configurations. 
As shown in Fig.~\ref{fig.9}, 
the achieved sum-rate of all schemes increases with $L$. 
This is because a larger number of RF chains provides more DoFs
for the design of the BS digital beamforming, 
enabling more flexible spatial multiplexing and improved interference management. 
As expected, 
the proposed ``RHS opt.'' scheme consistently outperforms the ``RHS fix.'' baseline 
across the considered range of $L$, 
which verifies the benefit of jointly optimizing the BS digital beamformer and the RHS holographic beamformer. 
Moreover, 
comparing the two antenna-spacing configurations, 
the $\frac{1}{4}\lambda$ spacing achieves a higher sum-rate than the $\frac{1}{2}\lambda$ spacing for all schemes. 
This observation confirms the advantage of deploying a holographic architecture with shorter spacing, 
which can offer a higher spatial resolution. 
Overall, Fig.~\ref{fig.9} demonstrates that increasing the number of RF chains 
and reducing the antenna spacing are both beneficial to the sum-rate performance, 
and the proposed joint design can better capitalize on these architectural advantages.

\section{Conclusions}

In this paper,
we propose a novel RHS-enabled SWIPT system.
By jointly optimizing the digital beamforming of the BS and the holographic beamforming of the RHS, 
we aim to maximize the sum-rate of all ID users while guaranteeing the individual EH requirements of each EH user.
Due to the non-convexity of the proposed optimization problem, 
we transform it into a convex one by jointly using the WMMSE and MM methods.
However, 
this solution has prohibitive computational complexity as the number of RF chains and/or RHS units increases.
To further reduce computational complexity, 
we develop a low-complexity algorithm based on the ADMM framework, 
which allows for the analytical update of all variables.
Simulation results demonstrate the advantages of the RHS-enabled BS compared with a conventional BS without RHS.
Moreover, 
compared with the solver-based solution, 
the low-complexity method significantly reduces the runtime without any performance degradation.

\appendix
\subsection{Invertibility and Numerical Stability of the Matrix Inversions}\label{app:matrix_inversion}
\normalem

{
In this subsection, 
we analyze the invertibility and numerical stability of the matrix inversion involved in the closed-form update of $\mathbf{w}_{IE}$. 
The matrix to be inverted is given by
$\bar{\mathbf{B}}_5
=
\mathbf{B}_5+\frac{\rho}{2}(G+1)\mathbf{I}_{L(K+G)}$.
Since $\mathbf{B}_5\succeq \mathbf{0}$ and $\rho>0$, for any nonzero vector $\mathbf{x}\in\mathbb{C}^{L(K+G)}$, we have
\begin{equation}
\mathbf{x}^H\bar{\mathbf{B}}_5\mathbf{x}
=
\mathbf{x}^H\mathbf{B}_5\mathbf{x}
+
\frac{\rho}{2}(G+1)\|\mathbf{x}\|^2
\geq
\frac{\rho}{2}(G+1)\|\mathbf{x}\|^2
>
0.
\end{equation}
Therefore, $\bar{\mathbf{B}}_5$ is Hermitian positive definite and hence nonsingular. 
Moreover, its eigenvalues satisfy
$\lambda_{\min}(\bar{\mathbf{B}}_5)
\geq
\frac{\rho}{2}(G+1)$,
and
$\lambda_{\max}(\bar{\mathbf{B}}_5)
\leq
\lambda_{\max}(\mathbf{B}_5)
+
\frac{\rho}{2}(G+1)$.

And, the condition number of $\bar{\mathbf{B}}_5$ is upper bounded by
\begin{equation}
\kappa(\bar{\mathbf{B}}_5)
\leq
\frac{
\lambda_{\max}(\mathbf{B}_5)+\frac{\rho}{2}(G+1)
}{
\frac{\rho}{2}(G+1)
}.
\end{equation}
This shows that the ADMM penalty parameter $\rho$ acts as a diagonal-loading term, 
which guarantees the invertibility of $\bar{\mathbf{B}}_5$ and improves numerical stability. 
}

\subsection{The Solution of (P8)}\label{app:P8_solution}
\normalem

{
First,
the Lagrangian function of (P8) is given as
\begin{align}
\mathcal{L}(\mathbf{f}_{0},\mu)
=&\Vert \mathbf{f}_{0} \Vert_2^2
-2\text{Re}\{ (\rho^{-1}\boldsymbol{\lambda}_0+\mathbf{w}_{IE})^H  \mathbf{f}_{0}\} \label{P8_1}\\
&+ \mu (\mathbf{f}_{0}^H\mathbf{f}_{0}- P_t), \nonumber
\end{align}
with the coefficient $\mu$ being the Lagrangian multiplier.

By taking the first-order derivative of $\mathcal{L}(\mathbf{f}_{0},\mu)$ w.r.t. $\mathbf{f}_{0}$ and setting it to zero, we obtain
\begin{align}
\frac{\partial \mathcal{L}(\mathbf{f}_{0},\mu)}{\partial \mathbf{f}_{0} } = \mathbf{0},
\end{align}
which yields
\begin{align}
\mathbf{f}_{0} = \frac{\rho^{-1}\boldsymbol{\lambda}_0+\mathbf{w}_{IE}}{ 1 + \mu }. \label{P8_closed_solution}
\end{align}

Substituting (\ref{P8_closed_solution}) into the power constraint (\ref{P8_c_1}) leads to
\begin{align}
\frac{\Vert\rho^{-1}\boldsymbol{\lambda}_0+\mathbf{w}_{IE} \Vert_2^2}{ (1+\mu )^2 }\leq P_t. \label{P8_proof_power}
\end{align}

Since the left-hand side is monotonically decreasing in $\mu$, 
the optimal solution of (P8) falls into two cases:
\begin{itemize}
\item[] \uwave{CASE-I}: If $\mu=0$ satisfies (\ref{P8_proof_power}), then
\begin{align}
\mathbf{f}_{0}^{\star} = \rho^{-1}\boldsymbol{\lambda}_0+\mathbf{w}_{IE}.
\end{align}
\item[] \uwave{CASE-II}: Otherwise, $\mu>0$ and the optimal solution is
\begin{align}
\mathbf{f}_{0}^{\star} = \sqrt{P_t}\frac{\rho^{-1}\boldsymbol{\lambda}_0+\mathbf{w}_{IE}}
{\Vert\rho^{-1}\boldsymbol{\lambda}_0+\mathbf{w}_{IE}\Vert_2}. \label{P8_2}
\end{align}
\end{itemize}
}



\end{document}